\documentstyle[11pt,aaspp4]{article}

\def\et{et al.}
\def\kms{km s$^{-1}$}
\def\ha{H$\alpha$}

\def\solar{\ifmmode_{\mathord\odot}\;\else$_{\mathord\odot}\;$\fi}
\def\ergsec{ergs s$^{-1}$}
\def\HII{H$\,${\sc ii}}
\def\HI{H$\,${\sc i}}

\def\lfir{L$_{FIR}$}
\def\ltir{L$_{TIR}$}
\def\fir{f$_{FIR}$}
\def\mum{$\mu$m}
\def\foi{f$_{[OI]}$}
\def\foiii{f$_{[OIII]}$}

\def\fcii{f$_{[CII]}$}
\def\fciifir{L$_{[CII]}$/L$_{FIR}$}
\def\lcii{L$_{[CII]}$}
\def\lha{L$_{H\alpha}$}
\def\fha{f$_{H\alpha}$}
\def\lco{L$_{CO}$}
\def\lb{L$_B$}
\def\firfb{L$_{FIR}$/L$_B$}
\def\flwtwo{f$_{6.75}$}
\def\flwthree{f$_{15}$}
\def\lwcolor{f$_{6.75}$/f$_{15}$}
\def\irascolor{f$_{60}$/f$_{100}$}

\def\lwtwoha{f$_{6.75}$/f$_{H\alpha}$}
\def\lwthreeha{f$_{15}$/f$_{H\alpha}$}
\def\ciilwtwo{f$_{[CII]}$/f$_{6.75}$}
\def\ciilwthree{f$_{[CII]}$/f$_{15}$}
\def\oioi{f$_{[OI]145}$/f$_{[OI]63}$}

\def\oiciipdr{L$_{[OI]}$/L$_{[CII],PDR}$}
\def\CII{C$\,${\sc ii}}
\def\OII{O$\,${\sc ii}}
\def\OI{O$\,${\sc i}}
\def\OIII{O$\,${\sc iii}}
\def\NII{N$\,${\sc ii}}
\def\NIII{N$\,${\sc iii}}

\begin{document}

\title{The Interstellar Medium of Star-Forming Irregular Galaxies: The
View with {\it ISO}\footnote{\rm Based
on observations made with
{\it ISO}, an ESA project with instruments funded by ESA Member
States (especially the PI countries: France,
Germany, the Netherlands, and the United Kingdom), and with the participation 
of ISAS and NASA.} 
}

\author{Deidre A.\ Hunter\footnote{\rm Visiting Astronomer, Cerro Tololo 
Interamerican
Observatory, National Optical Astronomy Observatory,
which is operated by the Association of Universities in Astronomy, Inc.,
under cooperative agreement with the National Science Foundation.}
}
\affil{Lowell Observatory, 1400 West Mars Hill Road, Flagstaff, Arizona 86001
USA;
\\dah@lowell.edu}

\author{Michael Kaufman}
\affil{Department of Physics, San Jose State University, San Jose, CA 95192 USA;
\\kaufman@ism.arc.nasa.gov}

\author{David J.\ Hollenbach, Robert H.\ Rubin}
\affil{NASA/Ames Research Center, MS 245-3, Moffett Field, CA 94035 USA;
\\ hollenbach@ism.arc.nasa.gov,
rubin@cygnus.arc.nasa.gov}

\author{Sangeeta Malhotra}
\affil{Johns Hopkins University, Charles and 34th Street, Bloomberg Center,
Baltimore, MD 21210 USA; san@tarkus.pha.jhu.edu}

\author{Daniel A.\ Dale, James R.\ Brauher, Nancy A.\ Silbermann, George Helou, 
Alessandra Contursi, and Steven D.\ Lord}
\affil{IPAC, 100-22, California Institute of Technology, Pasadena, CA 91125 USA;
\\jimby@ipac.caltech.edu, dad@ipac.caltech.edu,
nancys@ipac.caltech.edu, gxh@ipac.caltech.edu,
contursi@ipac.caltech.edu, lord@ipac.caltech.edu}

\begin{abstract}

We present mid-infrared imaging and far-infrared (FIR) spectroscopy
of 5 IBm galaxies observed by {\it ISO} as part
of our larger study of the interstellar medium of galaxies. 
Most of the irregulars in our sample are very
actively forming stars, and one is a starburst system.
Thus, most are not typical Im galaxies.
The mid-infrared imaging 
was in a band centered at 6.75 \mum\ that
is dominated by polycyclic aromatic hydrocarbons (PAHs) 
and in a band centered at 15 \mum\ that is dominated by
small dust grains. The spectroscopy of 3 of the galaxies 
includes [\CII]$\lambda$158 \mum\
and [\OI]$\lambda$63 \mum, important coolants of photodissociation regions
(PDRs),
and [\OIII]$\lambda$88 \mum\ and [\NII]$\lambda$122 \mum, which
come from ionized gas. [\OI]$\lambda$145 \mum\ and [\OIII]$\lambda$52 \mum\
were measured in one galaxy as well.
These data are combined with PDR and \HII\ region models to deduce
properties of the interstellar medium of these galaxies.

We find a decrease in PAH emission in our irregulars relative to small grain,
FIR, and \ha\ emissions for increasing FIR color temperature which 
we interpret
as an increase in the radiation field due to star formation 
resulting in a decrease in PAH emission.
The \lwthreeha\ ratio is constant for our irregulars, and
we suggest that the 15 \mum\ emission in these irregulars
is being generated by the transient heating of small dust grains by
single photon events, possibly
Ly$\alpha$ photons trapped in \HII\ regions. 
The low \lwthreeha\ ratio, as well as the high
\fcii/\flwthree\ ratio, in our irregulars compared to spirals
may be due to the 
lower overall dust content, resulting in fewer dust grains being, 
on average, near
heating sources. 

We find that, as in spirals, 
a large fraction of the [\CII] emission comes from PDRs.
This is partly a consequence of the high average stellar effective temperatures
in these irregulars.
However, our irregulars have high [\CII] emission relative to FIR, 
PAH, and small grain emission compared to spirals.
If the PAHs that produce
the 6.75 \mum\ emission and the PAHs that heat the PDR are the same,
then the much
higher \fcii/\flwtwo\ ratio in irregulars would require that
the PAHs in irregulars produce several times more heat than the PAHs in spirals.
Alternatively, 
the carrier of the 6.75 \mum\ feature tracks,
but contributes only a part of, the PDR heating, that is due mostly
to small grains or other PAHs. In that case, our irregulars 
would have a higher proportion of the PAHs that heat 
the PDRs compared to the PAHs that produce the 6.75 \mum\ feature.

The high f$_{[OIII]}$/f$_{[CII]}$ ratio
may indicate a smaller solid angle
of optically thick PDRs outside the \HII\ regions compared to spirals.
The very high L$_{[CII]}$/L$_{CO}$ ratios among our sample
of irregulars
could be accounted for
by a very thick [\CII]
shell around a tiny CO core in irregulars, and PDR models for one galaxy
are consistent with this.

The average densities of the PDRs and far-ultraviolet
stellar radiation fields hitting the PDRs are much
higher in two of our irregulars than in most normal
spirals; the third irregular has properties like those in typical
spirals.
We deduce the presence of several molecular clouds in each galaxy with
masses much larger than typical GMCs. 

\end{abstract}


\section{Introduction}

The infrared wavelength region offers a variety of diagnostics of the physical
state of the interstellar medium (ISM) in galaxies, and the {\it Infrared 
Space Observatory}
({\it ISO}, Kessler \et\ 1996) has enabled astronomers to observe many of these
infrared features
to unprecedented sensitivity levels. 
Under the US Guaranteed Time project to examine the ISM
of normal galaxies, we have used {\it ISO} to observe a suite of 
atomic and ionic fine structure lines and the infrared continuum of a large
sample of galaxies (Helou \et\ 1996). The 69 galaxies in our sample span the
full range of Hubble types from early-type E/S0 galaxies to 
late-type irregulars.
For a list of galaxies in the larger sample see Dale \et\ (2000).
In this paper we report on the group of irregular galaxies in our sample
and how they compare to our larger sample of spiral galaxies.
The irregulars are interesting 
because they are more similar to each other and more
different as a group from spirals
in, for example,
dust column depth and because they are the extreme end of 
the Hubble sequence of
disk galaxies in many galactic properties.
However, the reader should keep in mind that the irregulars in our
sample are not all typical of the Im class since most are very
actively forming stars and one is a starburst. The irregular
galaxies in our sample are described in detail in the next section.

The data obtained with {\it ISO} for our project include broad-band 
images of galaxies
at 6.75 \mum\ and 15 $\mu$m taken with CAM (Cesarsky \et\ 1996a) 
and fluxes of atomic and ionic fine-structure transitions measured
using the Long Wavelength Spectrometer (LWS, Clegg \et\ 1996).
The mid-infrared CAM images are in the regime where light from red
stars is decreasing with wavelength
and emission from ISM processes has become dominant.
The CAM images at 6.75 \mum\ (LW2, $\Delta\lambda$ of 3.5 \mum) include
major Polycyclic Aromatic Hydrocarbon (PAH) features at 6.2 \mum, 
7.7 \mum, and 8.6 \mum\ (Helou \et\ 2000).
The CAM images at 15 \mum\ (LW3, $\Delta\lambda$ of 6 \mum), 
on the other hand, are most sensitive to thermal emission due to
very small grains attaining high temperatures
(Dale \et\ 2000, Malhotra \et\ 2001).

The LWS was used in grating line-mode to obtain emission spectra 
of primary diagnostics of the ISM. These lines include
[\CII]$\lambda$158 \mum, [\NII]$\lambda$122 \mum, [\OIII]$\lambda$88 \mum, 
and [\OI]$\lambda$63 \mum. The [\CII] and [\OI]
are cooling lines for the
atomic gas and probe the conditions 
in photodissociation regions (PDRs), the warm neutral gas cloud surfaces,
which constitute a large fraction of the neutral medium in a galaxy.
The [\NII] and [\OIII] emission lines probe conditions in the \HII\ regions.
A subset of galaxies, including one irregular,
was also observed 
at [\OI]$\lambda$145 \mum\ and
[\OIII]$\lambda$52 \mum . 
The LWS beam of $\sim$75\arcsec\ diameter
included most of the optical galaxy for our distant sample of 60 galaxies and was
used to explore different environments within our nearby sample of 9 galaxies.
(The ``distant'' sample is defined as galaxies that are not well resolved with respect
to the size of the LWS aperture).

We supplement the {\it ISO} data with infrared continuum fluxes 
measured with the {\it Infrared Astronomical Satellite} ({\it IRAS})
at 12, 25, 60, and 100 \mum. We also have optical images that show us what
the optical counterparts are to the infrared features seen in the CAM images,
provide B-band luminosities for comparison to the infrared, and 
show us the amount and locations of current star formation in these galaxies.
Below we discuss what is known of the galaxies in our sample, the
observations that we use here, the results from the mid-infrared CAM 
imaging---both in terms of morphology and flux ratios, and the infrared
line ratios for the ionized and neutral gas and what they imply about
the physical conditions in the PDRs.

\section{The Galaxies}

The irregular galaxies in our sample include NGC 1156, NGC 1569, 
NGC 2366, NGC 6822, 
and IC 4662.
All of the galaxies that we discuss here are classed as IBm, barred
Magellanic-type irregulars, by de Vaucouleurs \et\ (1991; $\equiv$RC3).
Most of the irregulars in our sample are {\it not typical} of the Im class 
since the truly
typical Im galaxy is too faint in the far-infrared (FIR) for us to have
considered observing with {\it ISO}, particularly with the LWS
(Hunter \et\ 1989, Melisse \& Israel 1994). 
NGC 1156, NGC 1569, and IC 4662 are at the high end of the
distribution of surface brightness and star formation activity.
NGC 2366 is not, but it contains a supergiant \HII\ region.
NGC 6822 {\it is} a typical irregular and we did detect it 
(barely) with CAM but it is not bright enough in the
FIR for LWS observations within our time allocation.
The galaxies and pertinent global characteristics
are given in Table \ref{tabgal}. Optical \ha\ images
are shown in Figure \ref{figvha} with V-band image contours superposed.

At an M$_B$ of $-$18.0,
NGC 1156 is at the high end of the range of luminosities seen in normal,
non-interacting irregular galaxies (Hunter 1997), and is about 25\% brighter
than the Large Magellanic Cloud in the B band. 
Karachentsev, Musella, \& Grimaldi
(1996) describe NGC 1156 as ``the less disturbed galaxy in the Local
Universe,'' and there are no catalogued galaxy neighbors within 0.7 Mpc
and $\pm$150 \kms. NGC 1156 has a high rate of star formation
and \HII\ regions are crowded over the disk.

NGC 1569 
has the highest star formation rate of the sample of 69 irregulars
examined by Hunter (1997) as determined from its \ha\ luminosity. 
It has just undergone a true burst
of star formation (Gallagher,
Hunter, \& Tutukov 1984; Israel \& van Driel 1990; Vallenari \& Bomans 1996;
Greggio \et\ 1998), 
meaning that its
recent global star formation rate (SFR) is statistically significantly 
elevated compared
to its average past rate. In addition it contains two
luminous super star clusters that are likely young versions of
globular clusters (Arp \& Sandage 1985;
O'Connell, Gallagher, \& Hunter 1994),
and ionized gas filaments that extend well beyond the main optical
body of the galaxy (de Vaucouleurs, de Vaucouleurs, \& Pence 1974;
Hunter \& Gallagher 1990, 1992, 1997;
Hunter, Hawley, \& Gallagher 1993).
The starburst is said to have ended $\sim$4--10 Myrs ago but the presence
of \HII\ regions 
and 5 giant molecular clouds (GMCs)
(Taylor \et\ 1999) suggests that we can consider star formation
as on-going. Virial masses of the molecular clouds suggest that
the CO--H$_2$ conversion factor should be 6.6$\pm$1.5 times the
Galactic value (Taylor \et\ 1999).
Stil \& Israel (1998) have detected a $7\times10^6$ M\solar\ \HI\ cloud
at 5 kpc from NGC 1569 and the hint of a bridge in \HI\ connecting them.
The suggestion is that interaction with this cloud is responsible for the
recent starburst.

NGC 2366, located in the M81 Group of galaxies, 
is a lower luminosity irregular, having almost the same absolute 
B magnitude as the 
Small Magellanic Cloud. It contains the supergiant \HII\ complex NGC 2363
at the southwest end of the bar and labelled in Figure \ref{figvha}. NGC 2363
is nearly twice as bright in \ha\ as the 30 Doradus nebula in the 
Large Magellanic Cloud and contains a large fraction of the total
current star-formation activity of the galaxy (Aparicio \et\ 1995,
Drissen \et\ 2000).
However, there is also another large \HII\ complex near NGC 2363 and
numerous smaller \HII\ regions scattered along
the disk.

NGC 6822 is a low luminosity irregular in the Local Group.
It contains a modest population of small \HII\ regions (Killen \& Dufour 1982;
Hodge, Kennicutt, \& Lee 1988; Massey \et\ 1995), 
and a modest star formation activity (Hoessel \& Anderson 1986, Hodge
\et\ 1991, Gallart \et\ 1996c).
The structure and stellar content
of the galaxy have been discussed by Hodge (1977),
and the SFR over the past 1 Gyr has been found
to be approximately continuous (Marconi \et\ 1995, Gallart \et\ 1996b,
Cohen \& Blakeslee 1998).
NGC 6822 is one of the few irregulars in which a population of 
asymptotic giant branch stars has been measured (Gallart \et\ 1994).

IC 4662 is the lowest luminosity irregular galaxy in our sample.
Nevertheless, it has a SFR per unit area that
is only a factor of two lower than that of NGC 1569.
The physical characteristics of this galaxy have been studied by
Pastoriza \& Dottori (1981) and Heydari-Malayeri, Melnick, \& Martin 
(1990). Our optical images show a collection of stars with
associated \ha\ emission that appears to be detached from the
main body of the galaxy. It is located 1.5\arcmin, or
890 pc at IC 4662's distance, to the southeast from the center of the galaxy.
Whether this is a very tiny, but separate, companion galaxy,
or a very unusually placed OB association we cannot tell.
For this paper, we will refer to it as IC 4662-A, and it is labelled
as such in Figure \ref{figvha}.
In the optical IC 4662 does not look disturbed, as could be the case
if it were interacting with a companion, but its SFR is
unusually high.
In addition there are 3 tiny \HII\ regions located well away from the
main body of the galaxy and roughly halfway between IC 4662 and IC 4662-A.

In addition to these 5 irregular galaxies that form part of our
distant sample of galaxies, 
our team has observed IC 10
with {\it ISO} as part of our sample of nearby, resolved (with respect
to the LWS beam) galaxies. 
Three positions in IC 10 were
observed with the LWS. 
The {\it ISO} LWS observations of
IC 10 are the subject of another paper (Lord \et\ 2000), and the
CAM observations have been discussed by Dale \et\ (1999). 
However, we will
call upon those results for comparison here. 
IC 10 has a B-band luminosity that is comparable to that of NGC 2366 but
a current SFR per unit area that is 17 times higher.
Massey \& Johnson (1998) found that the density of evolved massive 
stars in the Wolf-Rayet phase is several times higher than that of the Large
Magellanic Cloud which is a considerably more luminous galaxy.
For this reason,
Massey \& Armandroff (1995) have suggested that IC 10 too is undergoing
a burst of star formation.
IC 10 is also known to be unusual in having \HI\ gas that extends about 7 times
the optical dimensions of the galaxy (Huchtmeier 1979).
It is now known that that extended gas contains a large \HI\ cloud
(Wilcots \& Miller 1998),
like that near NGC 1569, and which has been 
suggested is falling into the galaxy causing the current heightened
star formation activity (Sait\-o \et\ 1992).
The [\CII]158 emission in IC 10 has also been mapped 
from NASA's Kuiper Airborne Observatory by Madden \et\ (1997),
and [CI] at 492 GHz and CO rotational transitions have been observed
by Bolatto \et\ (2000).

All of the irregular galaxies in our sample are metal-poor compared to
spirals, a general characteristic
of Im galaxies. The oxygen abundances are given in Table \ref{tabgal}.
The oxygen abundances [12$+$log(O/H)] range from 8.06 to 8.39 for 
these galaxies. This corresponds to $Z$ of 0.003 to 0.006, which
are 20--40\% of solar.

\section{Observations}

\subsection{{\it ISO} Data}

The observations of irregular galaxies obtained with {\it ISO} by our team
are listed in Table \ref{tabisoobs}. All of the irregular galaxies have 
been imaged in the mid-infrared 
through the LW2 and LW3 filters of CAM; NGC 1569
has been observed with additional filters but those observations will
not be discussed here (see Lu \et\ 2000). 
The full galaxy sample in our program observed with CAM,
the nature of the observations, and the data reduction
are discussed by Dale \et\ (2000). The CAM frames of each galaxy
are combined to produce a final 9\arcsec\ resolution.
The field of view of NGC 1156, NGC 1569, and IC 4662 is 4.45\arcmin,
that of IC 10 is 7.25\arcmin, that of NGC 2366 is 8.2\arcmin,
and that of NGC 6822 is 12.65\arcmin.
The LW2 and LW3 
fluxes, \flwtwo\ and \flwthree, for the irregular galaxies discussed
here are given in Table \ref{tabcam}. Note that foreground stars
and, in the case of NGC 2366, a background galaxy have been removed
from the images prior to measuring the fluxes.

To compare the CAM images with \ha\ and V-band images we needed
to determine the coordinate system of all the images. The CAM
images were rotated according to the information in the headers so
that north would be up and east to the left. For ground-based images
we used stars identified on the
optical image to determine an astrometric solution for the optical
image. When we compared the LW2 and LW3 CAM images with the V-band images,
where there are stars in common, we found that the two were offset
with respect to each other. Since the pointing accuracy of the
CAM images is not expected to be better than 12\arcsec, we have
assumed that the coordinate systems of the CAM images needed minor 
corrections and offset the LW2 and LW3 images to match the V-band images using
the stars in common. The offsets were $\leq$10\arcsec.

The irregular galaxies have also been observed in FIR atomic
and ionic fine-structure lines. The lines that have been observed are listed
in Table \ref{tabisoobs} and the spectra are shown in Figure \ref{figlws}. 
The full sample of galaxies observed with
the LWS, details concerning the observations and data analysis, and the 
statistical
results for the full sample are given by Malhotra \et\ (2001).
The emission-line fluxes for the irregular galaxies are given
in Table \ref{tablws}. Upper limits are 3$\sigma$. 
The [\CII] fluxes of NGC 1156 and NGC 1569 have been corrected for 
overlying foreground [\CII] emission in the Milky Way using observations
of a nearby off-galaxy position. The corrections are of order 14\%.
The IC 10 emission-fluxes have been corrected for the fact that IC 10
is extended with respect to the LWS beam, and that affects the
derivation of the surface brightness. The correction factors we
used were
0.59 for [\CII]$\lambda$158,
0.68 for [\OIII]$\lambda$88, and 0.84 for [\OI]$\lambda$63.

When we compare the irregulars to the rest of the galaxies
in our distant galaxy sample, we will delete NGC 4418.
It is a Seyfert and does not represent a normal galaxy.
In plots we will label the distant galaxy sample minus the irregulars as
spirals.

\subsection{Ground-based and Other Data}

Integrated infrared fluxes at 12, 25, 60, and 100 \mum\ of these galaxies 
were measured with {\it IRAS}. Those fluxes and the FIR luminosity
\lfir\ integrated from 40--120 \mum\
are given in Table \ref{tabiras}. The FIR flux is determined
from \fir$=1.26\times10^{-14}(2.58f_{60} + f_{100})$
W m$^{-2}$, where $f_{60}$ and $f_{100}$ are
the fluxes at 60 and 100 \mum, respectively, in Jy (Helou \et\ 1988).
The total infrared flux, TIR, for 3--1100 \mum\ is estimated from 
{\it ISO} and {\it IRAS} data using
the empirical formulation of Dale \et\ (2001).

We also obtained optical ground-based images of the irregulars at the Perkins
1.8 m telescope at Lowell Observatory from 1992 to 1995 using various 
TI 800$\times$800 CCDs.
The \ha\ images were obtained through a filter centered at 6566 \AA\ with
a FWHM of 32 \AA. An off-band image was taken through a filter centered
at 6440 \AA\ with a FWHM of 95 \AA. 
The \ha\ image of NGC 2366 was constructed from a mosaic of observations
at multiple positions.
We observed IC 4662 at the Cerro Tololo Interamerican Observatory (CTIO)
on 1999 September 17. We used a SITe 2048$\times$2048 CCD on the 1.5 m
telescope with a 75 \AA\ FWHM filter centered at 6600 \AA. The off-band
filter was a broad-band R filter.  
NGC 6822 was observed in 1988 with a 0.2 m Takahashi telescope that was
bolted to the side of the Hall 1.1 m telescope at Lowell Observatory.
Those observations are reported by Gallagher \et\ (1991).

The off-band image was shifted,
scaled, and subtracted from the \ha\ image to produce an image containing
only emission from \HII\ regions. These images are used to trace the
current star formation activity. Irregular galaxies contain significantly less
dust than spirals.
Therefore,
we do not expect that the optical \ha\ images 
have missed significant amounts of star formation (Hunter \et\ 1989).

We used both spectral photometric standard
stars and \HII\ regions in nearby galaxies to calibrate the \ha\ flux.  
The \ha\ calibrations took into account 
the blueshift of the passband with temperatures below
20 C and the different transmissions of the filter at the redshift of the 
calibrating \HII\ region and the object.
We note that the \ha\ flux that we
measure for IC 4662 is 58\% of that measured by 
Heydari-Malayeri \et\ (1990). 

In IC 4662 there are three tiny \HII\ regions
to the extreme southeast of the galaxy, two close together and a third
more separated from them. The two close together have a combined 
reddening corrected
\ha\ luminosity of 5.2$\times10^{36}$ ergs s$^{-1}$, and the third
has a luminosity of 3.3$\times10^{36}$
ergs s$^{-1}$. Thus, these \HII\ regions are comparable to the Orion nebula.
IC 4662-A has an \ha\ luminosity of 2.6$\times10^{38}$
ergs s$^{-1}$, 1/22 that of IC 4662 itself and
consistent with a large OB association, 
$\sim$25 O7 stars.

To trace the stars, we use broad-band V images.
V-band images of NGC 1156, NGC 1569, and NGC 2366
were obtained by P.\ Massey with the Kitt Peak
National Observatory\footnote{\rm A division of the
National Optical Astronomy Observatory,
which is operated by the Association of Universities in Astronomy, Inc.,
under cooperative agreement with the National Science Foundation.}
4 m telescope and a SITe 2048$\times$2048 CCD
from 1997 to 1998.
NGC 6822 was observed by C.\ F.\ Claver with the 4 m telescope
and a SITe 2048$\times$2048 CCD at CTIO
in 1996. 
We obtained V-band images of IC 4662
during our observing run at CTIO in 1999 September. 

The abundances and reddening of the ionized gas were measured for individual
\HII\ regions in NGC 1156 and NGC 2366 using a long-slit spectrograph
mounted on the Perkins 1.8 m telescope. Details of the observations
and analysis are given by Hunter \& Hoffman (1999).

The \ha\ luminosities of the irregulars have been corrected for reddening
using total gas E(B$-$V)$_g$ given in Table \ref{tabgal}. These were
determined from Balmer decrements in emission-line spectra. 
The reddening correction uses the reddening curve of Cardelli \et\ (1989).
Most stars, however, will not be as extincted as those in \HII\ regions.
Therefore, 
integrated B-band luminosities, taken from RC3, are corrected
for reddening using the foreground E(B$-$V)$_f$ plus 0.05 magnitude
to account for internal reddening of the stars.

Star formation rates are derived from the \ha\ luminosities using
a Salpeter stellar initial mass function from 0.1 to 100 M\solar\
(Hunter \& Gallagher 1986). The star formation rates are normalized
to the area of the optical galaxy using $\pi$R$_{25}^2$, where R$_{25}$
in kpc is the radius at a B-band surface brightness of 25 magnitudes
arcsec$^{-2}$. 

In plots we will include spiral galaxies from the larger distant
galaxy sample for comparison. \ha\ fluxes for these galaxies have
also been measured from \ha\ images obtained at Lowell Observatory,
Palomar Observatory, and Kitt Peak National Observatory.
The \ha\ fluxes of the spirals are corrected for reddening assuming
a total E(B$-$V)$_g$$=$E(B$-$V)$_f$$+$0.8 mag, where E(B$-$V)$_f$
is the Milky Way reddening determined by Burstein \& Heiles (1984).
The 0.8 term is added to account on average for typical reddening
of \HII\ regions in spiral galaxies. Their \ha\ luminosities are
converted to normalized star formation rates in the same way as was
done for the irregulars.
Integrated B-band luminosities are taken from RC3 and corrected for reddening
using the foreground E(B$-$V)$_f$ plus 0.15 magnitude to account for
internal reddening of the stars.

\section{Mid-Infrared Imaging}

\subsection{Morphology}

In Figure \ref{figcam} we show the CAM LW2 images of the irregulars
in our sample
with \ha\ contours superposed.
One can see that to a large degree the \flwtwo\ emission traces \ha\ emission:
1) {\bf NGC 1156}: In NGC 1156 most of the \flwtwo\ emission is found in 
two blobs that
coincide with the brightest \ha\ peaks. There is also some diffuse
\flwtwo\ emission and it is outlined by a lower surface brightness
\ha\ contour. The \flwthree\ emission is found entirely from the
two brightest \ha\ peaks.
2) {\bf NGC 1569}: In NGC 1569 the \flwtwo\ and \flwthree\ emissions 
are found in
two concentrations at
the center of the galaxy where the \ha\ emission is also brightest.
In Figure \ref{figcam} the positions of the 5 GMCs
mapped by Taylor \et\ (1999) are marked along with the positions of
the two superstar clusters. 
The main peak of \flwtwo\ emission in NGC 1569, the northwestern
blob, is resolvable into
two peaks itself. The brightest peak to the northwest coincides with
the bright \HII\ region there and the 5 molecular clouds are
situated around the edges. The fainter peak to the southeast
sits immediately to the north of a fainter \ha\ peak and 
offset to the south of
super star cluster A (the northwestern one of the two).
A fainter peak to the northeast of this concentration is associated
with an \HII\ region that is slightly detached from this \HII\
complex. Another much fainter peak well detached to the west
is not obviously associated with any \ha\ feature.
The second brightest concentration of \flwtwo\ emission, the southeastern
blob, is primarily associated with the second brightest \HII\ concentration.
The faint peak of \flwtwo\ emission to the extreme southeast,
and just to the west of the object labeled as a star,
is also not obviously associated with any \ha\ feature in NGC 1569.
3) {\bf NGC 2366}: In NGC 2366 most of the mid-infrared emission 
coincides with the
supergiant \HII\ region NGC 2363 in the southwestern part of the galaxy.
There is a much fainter region of \flwtwo\ emission to the northeast of that
which does not obviously correspond to anything in particular in
NGC 2366 in the optical, but it is comparable in brightness to other
noise elsewhere in the image beyond the optical galaxy.
4) {\bf NGC 6822}: In NGC 6822 the mid-infrared emission is 
found in the direction of 5 \HII\ regions
which are identified in Figure \ref{figcam}. There is a bit
of diffuse emission towards the center of the galaxy but it
is of too low signal-to-noise to measure with any degree of
confidence. The fluxes given in Table \ref{tabcam} are the
combined fluxes of the \HII\ regions.
5) {\bf IC 4662}: In IC 4662 we find the \flwtwo\ emission primarily
in two bright concentrations 
with some low surface brightness
diffuse emission around them. These are located near the center
of the galaxy. 
Comparison with the \ha\ image 
suggests that these two concentrations
are also the brightest \HII\ complexes in the galaxy.
At 15 \mum\ we primarily see the brighter of these three regions.
We do not detect the small detached \HII\ regions to the southeast
or IC 4662-A, further to the south.

This theme of the coincidence of mid-infrared emission and
\HII\ regions extends to a large degree to IC 10 as well. There one sees a
very bright, amorphous blob in the southeast that coincides
with the brightest region of \ha. There are also several
smaller blobs but without a strong correlation between
\ha\ intensity and \flwtwo\ intensity, and there is some lower surface
brightness emission that is not obviously connected with
\ha\ emission at all. Then to the northwest of the
brightest region, the \ha\ and 6.75 \mum\ emission together trace
a partial shell of diameter 470 pc (for a distance to the galaxy of
1 Mpc).

Thus, in our sample of irregulars the mid-infrared emission that we measure
is associated with regions of star-formation. This may be an 
issue of detectability. Irregulars have less dust and, hence,
the emission over-all in the infrared is reduced relative to
spirals. Furthermore, it is the higher surface brightness regions
that will preferentially be detected and the fainter, lower
surface brightness emission will be lost. In a study of the
mid-infrared surface brightness of several nearby galaxies,
including IC 10, Dale \et\ (1999) found that variations in
dust column density are the primary drivers of infrared
surface brightness differences. Therefore, it is reasonable
to expect that \HII\ regions and associated gas clouds, which have higher columns
of gas and dust compared to the general ISM, will be the places
that are easiest to detect.

\subsection{Flux Ratios}

Dale \et\ (1999) show that the \lwcolor\ ratio is sensitive
to the intensity of heating of the ISM and, hence, is a function
of the mid-infrared surface brightness. In regions of intense
heating this ratio is $\leq$1, whereas in regions of less intense
heating, such as interarm regions in spirals, this ratio is $>$1.
Furthermore, the ratio \lwcolor\ decreases as the surface brightness
rises.
In the Milky Way, regions far from \HII\ regions have \lwcolor\
ratios of 1--1.8 (Cesarsky \et\ 1996b) whereas in or near \HII\
regions values of order 0.2 are seen
(see, for example, Cesarsky \et\ 1996b,c; Contursi \et\ 1998).
The decrease in the \lwcolor\ ratio with star formation activity
is most likely due to 
a combination of PAH destruction in high SFR
environments and the increasing contribution of very small
grain thermal emission in the LW3 passband.
The decrease in PAH features, however, does not necessarily mean physical 
destruction
of the PAH itself.
If, for example, only singly ionized PAHs produce the 6.75 \mum\ feature
(Hudgins \et\ 1997), a change in the ionization state of the PAH could
decrease the 6.75 \mum\ intensity. Similarly, coagulation of PAHs onto
larger grains in dense regions could suppress the fluorescence
that leads to the 6.75 \mum\ emission.

In Table \ref{tabcam} we give the integrated \lwcolor\ ratios for the 
irregulars as well as the values measured for individual \HII\ regions
in these galaxies.
One can see that the supergiant \HII\ complex NGC 2363 and, because it
is dominated by this \HII\ complex, its galaxy NGC 2366 have the
lowest \lwcolor\ ratios, a value of 0.1. 
Given the large number of hot O stars that have
recently formed in NGC 2363 it is not surprising that NGC 2363
has a very low value of \lwcolor. 
The rest of the galaxies, as well as the rest of the \HII\ regions,
have higher values of
\lwcolor, from 0.2--0.6.  
This range is significantly lower than the 
\lwcolor\ value of unity typical for quiescent environments,
and the values are typical of star-forming regions.
The variation in the ratio is presumably due to variations
in the intensity of the local stellar radiation field and dust content,
although hardness of the radiation field is also a possibility,
but it is clear that the mid-infrared emission is being
dominated by the dust in or near \HII\ regions in the irregulars.

In Figure \ref{figlwcolor} we plot the \lwcolor\ ratio against
the FIR color temperature ratio \irascolor , after the figure of Dale \et\
(2000), in order to compare the irregulars to the large sample of spirals. 
The dashed line marks a \lwcolor\ ratio of 1. 
First, we see that the irregulars in our sample
all lie below this line, with values
of \lwcolor$<$1, as we have
already discussed. However, it is also true that most of the galaxies
in the sample of spirals also have ratios $<$1. This is due to the
fact that the integrated mid-infrared fluxes are dominated by the higher surface
brightness regions in the beam, which have low values of \lwcolor\
(Dale \et\ 1999); the observations are not as sensitive to the lower
surface brightness emission that has a higher \lwcolor\ ratio.

Second, we see that, compared to the spirals, the high 
infrared-luminosity irregular galaxies in our sample
do not have unusual ratios of \lwcolor, but they do extend the
range to the lowest values; all but one of the irregulars and IC 10 have values
at the low end of the range. This is consistent with the
mid-infrared emission that we detect in the irregular galaxies
being more dominated by star-forming regions than is the case in spirals.
We speculate that irregular galaxies with lower star formation rates
than those in our sample would show more moderate values of this ratio
if they could be detected outside the brightest few \HII\ regions.
For example, NGC 6822 is a more typical irregular, but with {\it ISO}
we have only detected the brightest \HII\ regions.

Third, as Dale \et\ (2000) found,
we see that the \lwcolor\ ratio drops as the FIR
ratio \irascolor\ becomes larger, implying warmer large dust grains. 
One interpretation of this effect is, that as the dust radiating
in the FIR becomes warmer, the emission from very
small grains at 15 \mum\ increases more than the PAH features
at 6.75 \mum.
All populations of dust grains should be sensitive to
the intensity of the stellar radiation field, but the 15 \mum\
emission is most sensitive because it lies on the Wien side of the 
blackbody spectrum.  An alternate possibility is that stronger
fields destroy the carriers of the 6.75 \mum\ PAH features in irregulars,
and we discuss support for this hypothesis below.
The exception to this trend is NGC 6822 which has a low value
of \lwcolor\ and a low \irascolor . However, the CAM fluxes
in NGC 6822 are quite weak, and the mid-infrared fluxes
may be more uncertain than the formal error-bars suggest.

If the mid-infrared ratio \lwcolor\ is sensitive
to the stellar radiation field, particularly in and
around star-forming regions, one might expect a correlation
with the SFR per unit area in a galaxy.
That is, the more star formation packed into a given area on average,
the more intense the hot star radiation, and
the lower 
the ratio \lwcolor .
We plot \lwcolor\ against the SFR in Figure \ref{figlwcolor}, where
the SFR is a globally averaged value. 
However, there is no correlation.
In fact, four of the irregulars
themselves span a range of a factor of 50 in normalized SFR
and yet have very similar \lwcolor\ ratios.
Most likely, 
the \lwcolor\ ratio depends on
{\it very} local conditions (Contursi \et\ 2000).
The early type stars are only a  few parsecs away from the absorbing
dust for these ratios, whereas the SFR is averaged over kiloparsec
scales. In other words, the SFR per unit area depends on how many star-forming
clouds populate a square kiloparsec of galactic disk, whereas the
FIR dust temperatures and the \lwcolor\ ratios depend only on how
far away the exciting stars lie from the absorbing dust in an individual
cloud. There is, therefore, little correlation between the two. 

Since the mid-infrared emission spatially traces the \ha\ emission so
well in our irregulars, we have considered the ratio of the
mid-infrared fluxes to the \ha\ flux,
\lwtwoha\ and \lwthreeha, both for galaxies as a whole
and for individual \HII\ regions.
The integrated ratios are shown for our irregular galaxies, as well
as for the spirals, in Figure \ref{figlwhalpha}, and are listed in Table
\ref{tabcam}.
The \lwtwoha\ and \lwthreeha\ ratios depend on a proper correction
for extinction at \ha. We have used an average internal E(B$-$V)
for the spirals and others might make a different choice (see, for
example, Kennicutt 1983). Thus, the spiral \fha\ are uncertain
by factors of 2--3, but that is still small compared to the trend
seen in Figure \ref{figlwhalpha} and small compared to the difference
between the bulk of the spirals and the irregulars.
The irregular with the lowest and most unusual mid-infrared to \ha\ ratios 
is NGC 6822, but again we note the possibility of uncertainty in
the mid-infrared fluxes in this galaxy. 

We see that compared to the spirals, the irregulars in our sample
have less mid-infrared emission relative to their \ha\
emission, especially for regions of high FIR color temperature.
The \lwtwoha\ ratio for spirals, irregulars, and \HII\ regions
varies by a factor of 1000, and the irregulars have ratios at
the low end or below those of spirals.
There is a little less scatter among the entire sample
in \lwthreeha\ than in \lwtwoha.
According to Dale \et\ (1999), the surface brightness in the
mid-infrared is sensitive to both heating intensity and dust column density,
although the dust column density usually plays the larger role
in the surface brightness differences from galaxy to galaxy.
The irregulars generally have a lower dust content than spirals
(Hunter \et\ 1989) due to their lower metallicity.
So, dust column density is likely a factor in 
why the irregulars have lower ratios
with respect to \ha\ than the spirals.
However, we note that the PAH emission appears to be more sensitive to
this than the small dust grain emission since the irregulars differ from the
spirals more in \lwtwoha\ than in \lwthreeha .

Among the irregulars themselves, 
we see that, except for NGC 6822, the irregulars in our sample
with the smallest
\flwtwo\ emission per \ha\ emission also have the warmest dust
in the FIR
whereas there is no difference with \flwthree\ emission;
our irregulars have very similiar \lwthreeha\ ratios.
Since the \ha\ luminosities of the irregulars in our sample vary by
a factor of 10, the constancy of \lwthreeha\ suggests that
the emission from small grains is keeping pace with the \ha\ emission.
It seems unlikely that the dust column density is varying among
our irregulars in just the right way to compensate for variations
in \ha\ emission. 
(The L$_{FIR}$/L$_B$ ratio varies by only a factor of 3 among our sample
of irregulars).
Thus, the small grains appear to be reacting 
to the hot stars that are also ionizing the gas that
emits at \ha.
In other words, more small grains are heated as more hot stars are formed
(but see Bolatto \et\ [2000] who argue for the destruction of small dust
grains in low metallicity, high ultraviolet radiation environments). In
addition, they are heated in such a way that their 15 \mum\ flux
rises linearly with \ha  , which tracks the ionizing luminosity of the
exciting stars if the \HII\ regions are significantly ionization-bounded.
In the ionization-bounded regime, \ha\ essentially ``counts ionizing
photons'' from the stars, and therefore it appears that the 15 \mum\ flux
is similarly counting exciting photons from the stars.  This could occur
if the 15 \mum\ flux is dominated by the transient heating of small
grains by single photons, for example. If those photons were trapped
Ly$\alpha$ photons in the \HII\ regions, then the dust column density 
is not a factor since the photons scatter until absorbed by dust.

However, if this is the case,
then, since the 6.75 \mum\ flux from PAHs is excited by single photons
as well, the decline in the \lwcolor\ ratio with increasing \irascolor\ 
seen in  Figure \ref{figlwcolor} must be explained with a destruction
of the 6.75 \mum\ PAH carrier with increasing \irascolor\ in irregular 
galaxies.  This appears
consistent with the decline in \lwtwoha\ with increasing \irascolor .
While the \lwthreeha\ ratio is constant among the irregulars,
the \lwtwoha\ ratio varies by a factor of 6 if we exclude NGC 6822.
In other words, as the \ha\ emission increases, the emission by PAHs in the
LW2 passband does not keep pace
(see also the UV/{\it ISO} comparison of Boselli \et\ 1997).
For NGC 6822 we are not sufficiently confident of the mid-infrared
fluxes to argue that NGC 6822 is truly abnormal.
From studies of star-forming regions
in the Milky Way, we expect that PAHs are more
easily destroyed than small grains in the presence of intense stellar radiation
fields (Cesarsky \et\ 1996b,c; Contursi \et\ 1998) or they are charged
in such a way that they do not produce the 6.75 \mum\ feature.
Thus, we are most likely seeing an increase in star formation intensity 
causing a decrease in PAH 6.75 \mum\ emission. 
This is given additional credence by Figure \ref{figlwfir} which plots
the ratios of \flwtwo/\fir\ and \flwthree/\fir\ against \irascolor.
We see that as the FIR color temperature goes up, the ratio of
PAH to FIR emission drops, while the ratio of emission from small
dust grains to FIR remains more nearly constant, dropping only slightly.
(See Dale \et\ 2001 for a more thorough discussion of this relationship).

There is a different trend among the spirals. First, 
there is a scatter of a factor of 40 in \lwthreeha\  in the spirals 
that is much greater than in the irregulars.  Second, the \lwthreeha\ ratio
in the spirals increases with increasing FIR color 
temperature.  The higher metallicity and, therefore, higher
dust column densities in spirals may mean that the small dust 
particles responsible
for the 15 \mum\ flux are
on average closer to the OB stars 
and warmer than in irregulars. The 15 \mum\ flux in
spirals may not be dominated by single photon transient heating, but
by the steady emission of warm grains radiating on the Wien side of the
blackbody spectrum
(Dale \et\ 2001, Helou 2001).
The scatter is then caused by different density
\HII\ regions that give somewhat different temperature small grains
(even for the same \irascolor\ which indicates the temperature of
larger grains further away). The rise in \lwthreeha\ with \irascolor\ 
is caused by the increased temperature of the small grains as the
large grains get warmer.

In the right panels of Figure \ref{figlwhalpha} we plot
the mid-infrared to \ha\ flux ratios against the FIR to B-band
luminosity ratio. The more FIR emission
there is relative to optical emission (that is, the more cool, large-grain dust
emission there is) the higher the mid-infrared emission is relative
to \ha\ emission for both the PAHs and the small dust grains.
The irregulars in our sample follow both of these trends, 
but sit at the low \lfir/\lb\
end of the distribution although, as previously noted, their values
are more constant in \lwthreeha\ than in \lwtwoha\ for their
\lfir/\lb.
Thus, in these irregulars, the amount of cold dust emission is 
related to the emission from PAHs, which is consistent with the 
premise that dust column densities
are important in determining the PAH emission and, possibly, that
PAH emission originates in the cooler neutral gas (PDR) and dust
that dominates the FIR dust emission.
However, the amount of emission from small dust grains is also related
to the amount of total dust emission from larger grains as well
as would be expected for a continuous dust population
(Giard \et\ 1994).

Motivated by Helou \et's (2001) observation that the ratio
of [\CII]158 emission to the emission of PAHs 
in the LW2 passband is constant in galaxies,
we have plotted the ratio of the [\CII] emission to the mid-infrared
emission against the FIR ratio \irascolor\ and 
the ratio L$_{FIR}$/L$_B$ in Figure \ref{figlwcii}.
As Helou \et\ found, 
we see that the \ciilwtwo\ ratio 
varies by only a factor of 5  for the spiral galaxies.
We see that in our irregulars compared to spirals
the [\CII] emission is elevated relative to
that of the PAH emission by a factor of nearly 3, but that it
also stays quite constant as \irascolor\ changes. 

The modest scatter in \ciilwtwo\ for a large variation in \irascolor\
among spirals is interpreted by Helou \et\ (2001) as due to
the fact that either the same PAHs which heat the PDR gas also produce the
\flwtwo\ emission, or that the ratio of the 6.75 \mum\ carrier to the
grain or PAH population that dominates the PDR gas heating stays constant. 
In the irregulars in our sample
the former interpretation is hard to understand because
the ratio of PDR gas heating to 6.75 \mum\ emission is up by a factor of
3.  This implies that in irregulars the PAHs which produce the
6.75 \mum\ feature and do the heating are quite different from their
counterpart PAHs in spirals.
This could result from the different processing history 
and make-up of the ISM in irregulars compared to spirals.
In the latter interpretation, the increase
in the \fcii/f$_{6.75}$ ratio in irregulars could be caused by an increase
in the PDR heating population relative to the 6.75 \mum\ carrier population
(although some [\CII] emission comes
from \HII\ regions where PAHs may be destroyed).  
However, this seems inconsistent with observations
that show that the relative PAH and dust contents remain the same
throughout the Milky Way
(Giard \et\ 1994) although other studies conclude that PAH fractions
vary on small scales (see discussion by Helou \et\ 2001).

In Figure \ref{figlwcii} we see that for a given \irascolor\
the \ciilwthree\ ratio for our irregulars is higher than in spirals and that
there 
is a noisy trend such that the warmer the large dust grains
in the FIR, the less [\CII] emission per small grain emission.
Furthermore, there is no trend of the \ciilwtwo\ ratio with L$_{FIR}$/L$_B$
but a very noisy trend of \ciilwthree\ with L$_{FIR}$/L$_B$ 
(see also Malhotra \et\ 2001). 
The higher \ciilwthree\ at a given \irascolor\ may be caused by the lower
dust content in irregulars due to their lower metallicity, and that results
in the average radiating grain being further away from the OB stars
and, hence, cooler on average.  Perhaps as well, small dust grains subjected to
transient heating by single photons, as we have suggested from the
\lwthreeha\ ratio,
are more efficient at radiating at 15 \mum.
The trend with \irascolor\ suggests that [\CII] increases
more slowly with increasing radiation fields than does the 15 \mum\
emission. This is consistent with PDR models that show a weak dependence
of [\CII] with increasing UV fields (Hollenbach, Takahashi, \& Tielens
1991). Variations in the incident spectral energy distribution may play a
role in variations in the observed ratios as well (Dale \et\ 2001).

\section{Infrared Line Data}

Far-infrared emission lines observed with {\it ISO} are useful diagnostics
of the ISM of galaxies. [\CII]$\lambda$158 \mum\ and [\OI]$\lambda$63 \mum\
are important cooling lines of the neutral atomic medium. 
[\NII]$\lambda$122 \mum, [\OIII]$\lambda$88 \mum, [\NIII]$\lambda$57 \mum,
and [\OIII]$\lambda$52 \mum\ emission come from \HII\ regions.
The \HII\ regions contribute to the [\CII]$\lambda$158 \mum\ emission
as well. 
[\OIII]$\lambda$88 \mum\ and
[\OIII]$\lambda$52 \mum\ together provide a measure of the electron density $n_e$
of the \HII\ region, or more strictly, of the O$^{++}$ region. 
Here we examine the {\it ISO} emission line ratios observed for
the irregulars NGC 1156, NGC 1569, and IC 4662. The other irregulars 
included in this study, NGC 2366 and NGC 6822, 
were not observed with the LWS. 
The observations refer to integrated emission from a large fraction
of the optical galaxy, and thus include the full range of galactic
environments. Furthermore, the contributions of the different galactic
environments are
luminosity-weighted.

\subsection{The Ionized Gas}

\subsubsection{[\OIII]$\lambda$52,88 and $n_e$}

The 
f$_{[OIII]52}$/f$_{[OIII]88}$
ratio is sensitive to the electron
density of the O$^{++}$ regions of \HII\ regions. 
This ratio is given in Table \ref{tablws} 
for NGC 1569 (see also Malhotra \et\ 2001). 
It has also been measured from {\it ISO} observations
for IC 10. From the methodology of Rubin \et\ (1994) we find that
the O$^{++}$ regions covered by the LWS beam in NGC 1569 have an average $n_e$
of $\leq$300 cm$^{-3}$. The emission ratio places $n_e$ in the low density limit
and we can only place an upper limit on $n_e$.
This value of $n_e$ is comparable to the
values measured for two spirals in our sample (Malhotra \et\ 2001).
The low values of \HII\
region densities are also consistent with the low ($\leq$200 cm$^{-3}$)
values observed in giant \HII\ regions in irregulars from the ratios of optical
[\OII]$\lambda$3726,3729 lines (see, for example, Hunter \& Hoffman 1999).
However, the [\OII] lines measure the density of the O$^+$ regions
and are more appropriate for the C$^+$ portions of \HII\ regions
that contribute to the observed
[\CII]  emission.

\subsubsection{[\OIII]$\lambda$88 \protect\mum\ and T$_{eff}$}

The \lha/L$_{[OIII]88}$ ratio can be used to estimate the effective
temperature of the ionizing stars when the oxygen abundance of the 
gas is taken into account. We have measured this ratio for NGC 1569
and IC 4662, and the values are given in Table \ref{tablws}.
We have combined the observed ratios with
the models of Rubin (1985)
with a density of 100 cm$^{-3}$ and
scaled for the oxygen abundances given in Table \ref{tabgal}.
We use a density of 100 cm$^{-3}$ since in the previous section we found
$n_e$ to be in the low density limit ($n_e\leq$300 cm$^{-3}$).
Furthermore, in this regime \ha\ and [\OIII] emissions have the same
functional dependence on $n_e$, so the ratio \lha\/L$_{[OIII]}$
is not dependent on the density.
We examine the resulting T$_{eff}$ for three model choices
of the total nebular Lyman continuum photons per second. These models
are appropriate for \HII\ regions that are smaller than the 
supergiant regions.
Because the LWS
beam encompasses a large fraction of the galaxy, these are rough estimates
of the average stellar effective temperature.

We find that T$_{eff}$ in NGC 1569 is about 40,000 K (log g$=$4.0 model)
for all three models
with the uncertainty in the measured ratio producing an uncertainty 
of order 1000 K. 
This T$_{eff}$ corresponds to
an O6.5V star (Conti \& Underhill 1988). 
The total \ha\ luminosity of NGC 1569 would correspond to the equivalent
of 4500 O6.5V stars
(Panagia 1973). This large number of O stars is consistent with the recent
starburst in that galaxy. It has been suggested that the starburst
ended 5--10 Myrs ago (Greggio \et\ 1998), 4 Myrs ago
(Vallenari \& Bomans 1996), and 10 Myrs ago (Israel \& van Driel 1990).
The hydrogen-burning lifetime of an
O6.5 star is about 6.4 Myrs at a metallicity of 0.001 (Schaller \et\ 1992).
This lifetime is more consistent with the most recent estimates for the
end of the starburst; 10 Myrs would be too long ago.

The models yield an average T$_{eff}$ in
IC 4662 of somewhat hotter than about 45,000 K (log g$=$4.5 model)
unless the \HII\ regions are all Orion-like in size, which is unlikely.
The \lha/L$_{[OIII]}$ ratio yields a T$_{eff}$ that is hotter than
45,000 K even if the ratio is larger by 1$\sigma$. 
A T$_{eff}$ of 45,000 K corresponds to an O4V star (Conti \& Underhill 1988).
The total \ha\ luminosity of IC 4662, however, would
correspond to only 50 equivalent O4V stars (or $\sim$400 O6.5V stars). 
Thus, the total number of
O stars in IC 4662 is significantly lower than in NGC 1569, but we
are catching those regions at a much younger age unless the stellar
initial mass function is highly top heavy which seems unlikely
(Massey \& Hunter 1998). The hydrogen-burning
lifetime of an O4 star is about 3 Myrs (Schaller \et\ 1992).

In both galaxies the average T$_{eff}$ are hotter than what 
are usually determined for
other galaxies. For example, T$_{eff}$ determined from
an [\OIII]88 measurement 
in the starburst nucleus of NGC 253 gives $\geq$34,500 K, which is
equivalent to an O8.5V star (Carral \et\ 1994).
For a zero-age main sequence stellar population of the same stellar
initial mass function, a higher T$_{eff}$ could result if individual
star-forming regions are richer in stars, rich enough to allow the rarer
highest mass stars to form (Massey \& Hunter 1998). In that case, then,
we would expect to see supergiant \HII\ regions associated with
these large concentrations of massive stars or perhaps super star clusters. The
luminosity function of \HII\ regions in NGC 1569 does show
that the \HII\ region population in NGC 1569 extends to 
luminous \HII\ regions (\lha$\sim10^{40}$ \ergsec)
(Youngblood \& Hunter 1999; we do not have an \HII\
region luminosity function for IC 4662), and NGC 1569 does contain
two super star clusters.
In addition, since NGC 1569
is a starburst system and, hence, the stellar population is coeval
to a greater extent than in most galaxies, the average age of the O-star
population is expected to be low compared to spirals with more continuous star
formation.

\subsubsection{[\OIII]$\lambda$88 and [\CII]}

The \foiii/\fcii\ and \foiii/\foi\ ratios are much higher 
in the irregulars
than they are in spirals. This is shown for \foiii/\fcii\ in 
Figure \ref{figoiiicii}. Plots of \foiii/\foi\ look similar to this
since [\OI] and [\CII] both come mostly from PDRs. 
We see that the irregulars have \foiii/\fcii\ ratios that are 
as much as  16 times higher than those in most spirals.
We use the
[\NII] observations in \S5.3.1 to show that most of the [\CII]
originates in PDRs as opposed to \HII\ gas whereas [\OIII]
comes from \HII\ regions.


Even if there were no PDR contribution to the [\CII] in irregulars, 
we would not expect differences of order 16 from spirals if the
effective temperatures of the stars were not considerably higher in
irregulars compared with spirals.  In spirals, the \HII\ regions contribute
of order 0.2 to 0.5 of the [\CII] emission (Malhotra et al 2001). 
Therefore, eliminating the PDR contribution would only increase the
\foiii/\fcii\ ratio by factors of 2-5 in spirals.  The irregulars 
can only gain the large factor of 16 by also having less [\CII]
compared with [\OIII] in the \HII\ region.  This is accomplished by
having higher effective temperature stars, which doubly ionize
most species and leave only small quantities of C$^+$ in the \HII\ regions.

However, if \HII\ regions in the irregulars in our sample were 
surrounded by optically thick PDRs (A$_V > 1$),
the \foiii/\fcii\ ratio would be much smaller than observed,
because such PDRs produce copious emission in [\CII], quite independent
of metallicity (Kaufman \et\ 1999). Therefore, the extremely
high ratios of \foiii/\fcii\ observed in the irregulars in our sample
can only be
achieved by having both high effective temperature stars and small
covering factors 
of optically thick PDRs outside the \HII\ regions
compared with normal spirals.
  

\subsection{The PDR and Neutral Gas}

\subsubsection{The [\CII]-to-CO Ratio}

Integrated \lcii/\lco(1--0) ratios have been shown to be related to the
global star formation activity of a galaxy (see, for example,
Pierini \et\ 1999).
Stacey \et\ (1991) have
shown that galaxies with starburst nuclei have
ratios comparable to those seen in Galactic star-forming regions,
while more quiescent galactic nuclei have ratios that are about
3 times smaller and comparable to values averaged over entire  Milky
Way GMCs. 
Stacey \et\ measured \lcii/\lco\ values of 900--6100 for normal spirals
and values of 1000--1800 for GMCs.
By contrast Stacey \et\ measured a ratio of 6100 for the starburst
galaxy M82 and values of 5900--14000 for Milky Way \HII\ regions. 
Poglitsch \et\ (1995) measure a ratio of 60000 in the supergiant
\HII\ complex 30 Doradus, and
Madden \et\ (1997) find values of 1500--10000 throughout
IC 10.
Smith \& Madden (1997) 
measured ratios of 14000 and 16000 for two
Virgo spirals. 
There are, however, also variations among similar clouds.
Israel \et\ (1996) report variations of up to a factor
of 100 among individual clouds in the LMC; they measured
ratios of 400--34000 in 4 clouds, and attribute the
variations to evolutionary effects.
At the other extreme, Malhotra \et\ (2000) extend the \lcii/\lco\
range to low values with {\it ISO} observations of early-type galaxies
where presumably the environment is quite quiescent.
Generally, regions of intense
star formation, including kpc-sized regions in galaxies,
have much higher \lcii/\lco\ compared to more quiescent regions.

To place our irregular galaxies in this context, we have examined the
\lcii/\lco\ ratios for NGC 1156, NGC 1569, and IC 4662, and they
are given in 
Table \ref{tablws}. The $^{12}$CO(1--0) observations are taken from
Hunter \& Sage (1993), Young \et\ (1995), and Heydari-Malayeri \et\ (1990),
and \lco$=0.119 D^2 S_{CO}$ from Kenney \& Young (1989) where
$S_{CO}$ is the CO flux in Jy \kms\ and D is the distance to the galaxy
in Mpc.
The lower limits on \lcii/\lco\ for NGC 1156 and IC 4662 result from
upper limits on \lco. The observed ratios in the irregulars are very
high: at least several times $10^4$. Galactic values of this level are also
seen in the irregulars
IC 10 (Madden \et\ 1997) and the LMC (Mochizuki \et\ 1994).

These values are much higher than
values Stacey \et\ (1991) and others measured for spiral 
nuclei and giant
molecular clouds. But, they are also higher 
than values measured for all but one of the Milky Way \HII\ regions and 
higher than that measured for the starburst galaxy M82.
This can also be seen graphically in Figure 2(b) of Pierini \et\ (1999)
where they plot \lcii/\lco\ against \lcii/\lfir. All of the galaxies 
in their sample and our irregulars have similar \lcii/\lfir, but our
irregulars sit at higher \lcii/\lco\ than all of the extragalactic
and Galactic sources.
The ratios for our three irregulars are not
corrected for the smaller beam size of the CO observations, although
the ratios of the beam sizes are given in Table \ref{tablws}, and in two
cases the ratios are lower limits because of upper limits on the CO flux.
Thus, the \lcii/\lco\ ratios of these irregulars could be even higher.

Pierini \et\ (1999) show
that there is a correlation between \lcii/\lco\ and \ha\ equivalent
width in the sense that galaxies with high equivalent widths, indicating
a higher relative star formation rate, have higher \lcii/\lco.
In our sample NGC 1569 has just undergone a starburst, and NGC 1156
and IC 4662 relatively high SFRs.
In NGC 1569 the starburst has blown a hole in the \HI\ (Israel \& van Driel 1990)
and severely disrupted the ISM. Yet in spite of this, NGC 1569's 
\lcii/\lco\ ratio is of the same order of magnitude as that of the other two
irregulars. Perhaps the five GMCs detected 
in NGC 1569 (Taylor \et\ 1999) are dominating both the [\CII] and CO emission.

The high \lcii/\lco\ ratio in the irregulars
could be an indication of a different geometry
of clouds in irregulars compared to spirals (e.g., Pak \et\ 1998,
Kaufman \et\ 1999, and references therein). If the PDR is a thin
skin around a CO core, both the [\CII] and CO luminosity from a cloud
go as R$^2$ and so the ratio is approximately constant for a dense
warm PDR layer. 
However, if the CO core is small and the [\CII] layer around the outside
is very thick, the CO luminosity goes as R$^2$ but the [\CII] goes
as (R$+\Delta$R)$^2$ and the \lcii/\lco\ ratio will be higher.
Thus, Im galaxies could have very thick PDR regions around tiny molecular
cores. 
This would also be consistent with the picture in which 
the molecular clouds are the dense cores of large, massive
atomic gas complexes (Rubio \et\ 1991).
In fact, below, we derive physical conditions of the PDRs from
models, and find that for NGC 1569 the PDR is 6 pc thick, thicker
than is usual in spirals.

It should also be kept in mind that \lco\ in irregulars
is probably an underestimate of the amount of molecular hydrogen
that is present compared to the case in higher metallicity spirals.
There is strong evidence that the 
conversion of 
I$_{CO}$ to N(H$_2$) depends on the metallicity  and that the CO
component of a cloud in metal poor irregulars is a much smaller
fraction of the H$_2$ cloud
(see, for example, Maloney \& 
Black 1988; Dettmar \& Heithausen 1989; 
Stacey 1993; Mochizuki \et\ 1994; Poglitsch \et\ 1995;
Verter \& Hodge 1995; 
Wilson 1995; Madden \et\ 1997; Smith \& Madden 1997;
but see also Wilson \& Reid 1991).
In the LMC where the metallicity is about half solar, 
virial masses of molecular clouds yield a correction
to the standard ratio of a factor of 6 (Cohen \et\ 1988).
A factor of 6.6 was also found for giant molecular clouds in NGC 1569
(Taylor \et\ 1999), but the 
correction factor to L$_{CO}$ remains somewhat controversial.

%

\subsubsection{[\CII]$\lambda$158 and FIR Emission}

Malhotra \et\ (1997, 2001) report a correlation between integrated \fciifir\
and \irascolor\ and \firfb\ for the distant galaxy sample. 
The bulk of the galaxies have similiar \fciifir\ ratios, but then as
the FIR dust temperature increases, the \fciifir\ ratio decreases
and \firfb\ increases.
For most galaxies L$_{[CII]}$/L$_{FIR}$ is about the same because
$G_0\propto n^{1.4}$.
Here
$G_0$ is the far-ultraviolet (FUV) stellar radiation flux,
in units of the value measured for the solar
neighborhood,
and $n$ is the neutral 
gas density.
Therefore, $G_0$/$n$ rises slowly with $G_0$, which
keeps the grain charge fixed and the gas heating efficiency via the grain
and PAH photoelectric heating mechanism fixed.
This implies similarities in star-forming clouds in different galaxies.
Malhotra \et\ (2001) interpret the decrease in \fciifir\ at high
values of \irascolor\ (FIR dust temperatures) in some galaxies
as due to less efficient gas heating
resulting from charged dust grains in high $G_0$/$n$ regimes
(Bakes \& Tielens 1994).
However, one should also question the normalization of \fciifir\
given the role of PAHs (Helou \et\ 2001).

Figure \ref{figciifir} shows the irregular galaxies in our sample
relative to the spirals
in L$_{[CII]}$/L$_{FIR}$. 
We see that NGC 1156 and a position in IC 10 have
among the highest
\fciifir\ ratios of the entire sample of galaxies. The other two irregulars
and other positions in IC 10 have \fciifir\ ratios that are comparable
to that of the bulk of the spirals. 
Furthermore, for a given FIR dust temperature, the irregular 
galaxies are seen to have a higher value of \fciifir\ than irregulars.

For comparison, Israel \et\ (1996) report a \fciifir\ ratio of
0.01 for the supergiant \HII\ complex 30 Doradus in the LMC
and, presumably because it is dominated by 30 Doradus, for the LMC
as a whole.
The ratio observed in 30 Doradus is, therefore, 2--4 times higher than
the integrated values for the irregulars in this sample. Even though
NGC 1569 is a starburst galaxy, it is significantly lower in
\fciifir\ compared to a supergiant star-forming region, implying
that
average conditions in NGC 1569 today are not too extreme.
This could
mean that the FIR emission in NGC 1569 is being dominated by
the remaining GMCs and is not reflecting
the disrupted nature of the rest of the ISM of the galaxy.
The LMC, on the other hand, has a much higher integrated value of \fciifir\ 
than NGC 1569
even though NGC 1569 has a recent SFR per unit area that
is 30 times higher. This suggests that a supergiant \HII\ region,
when it is present, can dominate the integrated properties of a galaxy.
We would predict, therefore, that NGC 2366 should have a \fciifir\
ratio that is like that of the LMC since it is dominated by the supergiant
\HII\ complex NGC 2363. Unfortunately, we do not have LWS observations
of NGC 2366.

The lower ratio of \fciifir\ in spirals compared to our irregulars for
the same \irascolor,
might be an indication that cooler sources of radiation (than hot OB
stars) may be contributing to the FIR luminosity
in those sources with low \fciifir . Another possibility is that
the ratio of $G_0/n$ is somewhat higher in those sources with low \fciifir,
resulting in more highly charged grains and less efficient gas heating.

In Figure \ref{figciifir} we also plot \fciifir\ against the global 
SFR per unit area for the irregulars and other galaxies in the distant
sample. We see that the bulk of the sample has a similar \fciifir\ ratio
with some decrease or more scatter of \fciifir\ as the SFR per 
unit area decreases for the spirals (the irregulars appear to go
counter to this trend, but the statistics are poor).
This trend is similar to what is seen by Pierini \et\ (1999) 
for a sample of Virgo
cluster spirals, where they see a plateau value of 0.004 in \fciifir\
and a drop to lower values for lower \ha\ equivalent widths.
They suggest that this plateau is an average of compact
and diffuse regions in the galaxy and that
diffuse regions begin to dominate as the SFR declines.
However, PDR models (Kaufman \et\ 1999)
and observations of diffuse regions in the Galaxy
show that diffuse regions have lower $G_0$/n and, hence, higher
\fciifir.
The near constancy with SFR means that a higher level of star formation
in a given area does not imply a higher local $G_0$/$n$ ratio.
Although $G_0$ goes
up in regions of star formation, Malhotra \et\ (2001) argue
that $n$ goes up too, so that ordinarily the average $G_0$/$n$,
and hence \fciifir, 
rises slowly.
However, individual star-forming regions may have enhanced
$G_0$/$n$ ratios, which lead to  inefficient heating by grains
and lower \fciifir.  The large scatter for low values of
SFR may be caused by a small number of star-forming regions
dominating this regime, so that the local conditions in a particular
star-forming region become more evident.

\subsubsection{[\OI]$\lambda$145,63}

The ratio \oioi\ is a measure of the temperature of the gas
or an indicator of optical depth in the 63 \mum\ line.
Malhotra \et\ (2001) interpret the increase of \oioi\ with increasing
\firfb\ in the large sample of galaxies
as an indication of optical depth effects in [\OI]63.
We have \oioi\ for NGC 1569 and that measurement is shown in Figure
\ref{figoioi}. We see that NGC 1569 falls at the tail end of the relationship
traced by the spirals. Thus, [\OI]63 in NGC 1569 appears to be affected
by optical depth as in spiral galaxies, but NGC 1569 lies
at the low optical depth end of the distribution.

\subsection{Physical Conditions in the PDRs in Irregulars}

The FIR emission-line ratios can be combined with PDR models
in order to derive fundamental parameters 
that describe the state
of the ISM in the irregular galaxies in our sample
(Kaufman \et\ 1999). In particular,
the FIR line ratios yield information on the far-ultraviolet stellar 
radiation field, $G_0$,
and the neutral gas density, $n$, of H nuclei in PDRs.

\subsubsection{The PDR and \protect\HII\ Region Contributions to [\CII]}

A primary diagnostic is the [\CII] line, but both
PDRs and dense and diffuse \HII\ regions can contribute to this line flux.
In Figure \ref{figciiha} we plot the \lcii/\lha\ ratio against
global L$_{FIR}$/L$_B$. 
The correlation of \lcii/\lha\ with \lfir/\L$_B$ resembles that of
\lwtwoha\ and \lwthreeha\ with \lfir/L$_B$ shown in Figure \ref{figlwhalpha}
although there is more scatter in the \lcii/\lha\ plot.
As cold dust emission increases relative to the optical, the [\CII],
PAH, and small grain emissions all increase to some extent relative to \ha.
Thus, the amount of [\CII] emisison is related to the amount of
cold dust emission, as expected.
We see that the three
irregulars in our sample
have the lowest \lcii/\lha\ ratio of the distant
galaxy sample. 
However, \HII\ regions could be a significant source of [\CII] emission,
and that contribution needs to be removed in order to examine the 
emission from PDRs.

We have, therefore, determined corrections to \lcii\ 
for the \HII\ contribution following
the procedure of Malhotra \et\ (2001).
They used the fact that [\NII]$\lambda$122
\mum\ emission comes only from the dense and diffuse ionized gas (not the
PDR gas) and derive the ionized gas 
contribution to [\CII] as 4.5 times the flux of [\NII].
This correction factor assumes 
a C/N ratio of 1.9
and that may not be accurate for metal-poor irregulars.
If irregular galaxies such as those for which we have LWS spectra
have higher C/N ratios, as indicated by Garnett \et\ (1995),
the correction to [\CII] would be correspondingly higher.
On the other hand, for the three
irregulars we only have upper limits to L$_{[NII]}$
and hence lower limits to the PDR contribution to the [\CII] flux,
L$_{[CII],PDR}$. 
At least 61\% of the [\CII] emission in the irregulars is from PDRs.
The L$_{[OI]}$/L$_{[CII],PDR}$ ratios 
are given in Table \ref{tablws} as a range of values with the lower
end of the range determined assuming that all of [\CII] comes from the PDRs
and the upper end of the range determined assuming a correction to [\CII]
for \HII\ regions that assumes the upper limit of the [\NII] flux.

For IC 10 the contribution of ionized gas to [\CII] was found to
be $\sim$20\% by Madden \et\ (1997).
Stacey \et\ (1993) also found that 25\%--30\% of [\CII] emission
in the spirals NGC 6946 and NGC 891 comes from diffuse ionized gas,
and Malhotra \et\ (2001) find that the correction is typically 30\%.
Thus, unless we are underestimating the amount of [\CII] coming
from \HII\ regions because of a different C/N ratio,
at most the irregulars in our sample are similar to other galaxies in 
proportions of [\CII] coming from PDR and ionized gas, and potentially
our irregulars have a somewhat lower proportion coming from ionized gas.
This is a surprise because other parameters have led us to
conclude that the integrated properties of
irregular galaxies are dominated by \HII\
regions more so than the spirals.  However, because of the high
effective temperature of the exciting stars,
the \HII\ regions in these irregulars
will be in high ionization states, resulting in 
less C$^+$ in the \HII\ gas than would otherwise be expected.

The ratios of [\OI] to [\CII] are shown in Figure \ref{figoicii}
for the corrected and uncorrected [\CII] flux. The points for the
3 irregulars are upper limits for \oiciipdr,
but those for two
regions in IC 10 are measurements since the [\NII] line was detected there.
We see that our irregulars have \oiciipdr\ ratios that are 
lower for a given \irascolor\ ratio compared to spirals.

\subsubsection{Results from PDR Models}

We have used the LWS line strengths in conjunction with 
models that are slightly modified from those presented by Kaufman \et\ (1999).
The analysis by Kaufman \et\ is most appropriate for individual
nearby molecular clouds where
FUV illumination occurs primarily from one side
of a molecular cloud.
The models have been modified to consider integrated emission from
a large fraction of a galaxy by assuming that the interstellar clouds
can be illuminated from any side and that optically thin [\CII] and FIR
continuum reach the observer from any side.
Because it is, however, optically thick, [\OI] emission 
is assumed to come only from
the side closest to the observer.
The ratio (\lcii$+$L$_{[OI]}$)/\ltir\ measures the
efficiency with which FUV photons are converted
to emission lines. This is combined with the \oiciipdr\ ratio to derive
$G_0$, $n$, and the average gas surface temperature T$_{gas}$ and pressure $P$
of the interstellar gas producing the line and
the FIR continuum in PDRs.

In this manner, we derive $G_0$, $n$, T$_{gas}$, and $P$
given in Table \ref{tabmodel} for PDRs in NGC 1156, NGC 1569, and IC 4662.
We see that $G_0$ is $1.1\times10^3$--$2.4\times10^4$, $n$ is 
$1.3\times10^3$--$2.8\times10^4$ cm$^{-3}$, and T$_{gas}$
is 310--725 K.
The conditions in IC 4662 seem to be the most severe of the three
irregulars.

For the larger group of spiral galaxies, Malhotra \et\ (2001) derived
$G_0$ of 10$^2$--10$^{4.5}$ and $n$ of 10$^2$--10$^{4.5}$ cm$^{-3}$.
A comparison of the irregulars with Figure 9 of Malhotra \et\ (2001)
shows how $G_0$ in the irregulars compares with typical values in
spirals: NGC 1156 has a $G_0$ that is comparable to the bulk of spirals,
IC 4662's $G_0$ is higher than those of all but a few spirals,
and NGC 1569 has a $G_0$ that is higher than those of most of the spirals.
The values of $G_0$ and $n$ derived for the irregulars,
especially NGC 1569 and IC 4662, are similar to the values derived
for the peculiar galaxy M82. 
For M82 Colbert \et\ (1999) analyze LWS spectra to
find that $G_0$ is 630 and $n$ is
2000 cm$^{-3}$, but
Kaufman \et\ (1999) find a higher $G_0$ of 10$^{3.5}$
and a higher $n$ of 10$^4$ cm$^{-3}$ for the center of M82
from analysis of Kuiper Airborne
Observatory data (Lord \et\ 1996).
Also, the central regions of IC 10, a galaxy that has many
similarities to NGC 1569, have $G_0$ that are
a factor of 6 lower, 300--500,
with densities that are similar, 
10$^3$--10$^4$ cm$^{-3}$ (Madden \et\ 1997).
However, in the starburst nuclei of two spirals, Carral \et\ (1994)
measure $G_0$, 10$^3$--10$^4$ that are similar to the three irregulars.
Thus, the illuminated neutral clouds in 
irregulars are within the range of values observed for
spiral galaxies but are at the high end, having a more intense
stellar radiation field reaching the clouds and a higher density than most.
Malhotra \et\ interpret the high T$_{gas}$ and
$P$ they derive for the spiral sample as indicating that most 
of the grain and gas heating is occuring very close to the concentrations
of young, hot stars in these GMCs, and this must be the case
in the irregulars as well.

We derive pressures $P$ in the PDRs of 4--200$\times10^5$ K cm$^{-3}$.
Elmegreen \& Hunter (2000) have estimated pressures of \HII\ regions
in a sample of 6 irregular galaxies (unfortunately, none of the galaxies
are in common with our LWS observations). They find \HII\ region
pressures of 10$^4$--10$^5$ K cm$^{-3}$. The pressures estimated
from \HII\ regions are therefore
lower than those in the PDRs of our galaxies, although both
pressure determinations are
uncertain. The \HII\ region pressures measured by Elmegreen and Hunter
are lower probably because these regions are evolved enough that they
no longer have PDRs around them.

The differences in physical PDR parameters between the irregulars
NGC 1569 and IC 4662
and normal
spiral galaxies imply that the star-forming clouds
in NGC 1569 and IC 4662 are different.
This is perhaps not a surprise since at least one of the galaxies is a
starburst. 
NGC 1569 is a starburst galaxy with a high population of massive
stars providing an intense FUV field, and
the ISM has been disrupted in many ways.  
The very high $G_0$ and $n$ of IC 4662 suggest that maybe this galaxy
has experienced some recent peculiar event like that of NGC 1569,
these properties are also consistent with the higher average stellar
T$_{eff}$ found in \S5.1.2.
NGC 1156, on the
other hand, is a normal, albeit active, irregular and its cloud properties
are correspondingly more normal.

In order to further investigate the properties of the ISM in these three 
irregular galaxies, we use the technique of Wolfire, Tielens \& Hollenbach
(1990) along with the model results of Kaufman \et\ (1999) modified for 
application to extragalactic sources. 
Given the PDR 
parameters derived from [\OI], [\CII] and FIR observations, the PDR models 
make specific predictions for the [\CII] cooling rate per atom and the [\CII]
intensity. By comparing the cooling rate per atom with the observed [\CII] flux
from PDRs, L$_{[CII],PDR}$, an estimate of the atomic gas mass 
M$_a$ in PDRs can be found.
Following the measurements of Taylor et al. (1999) for NGC1569, we assume a 
carbon abundance of 1/6 solar, so $x_C=2.3\times 10^{-5}$ in all three galaxies. 
From this, we find the mass of atomic gas on the outsides 
of molecular clouds in these three galaxies varies from 
$\sim 1.3\times 10^5 - 4\times 10^7\,M_{\sun}$; values are given in Table
\ref{tabmodel}. An estimate of the area filling
factor of PDR gas in a galaxy's central region producing [\CII] emission 
can be made by comparing the observed [\CII] intensity for this region with the
predicted [\CII] intensity from the PDR models. In order to make this 
comparison, we assume that the [\CII] emitting region lies within the 
contours shown on the 6.75 \mum\ images in Figure \ref{figcam} 
(see figure caption for \ha\ surface
brightness levels). Estimates of the 
PDR area filling factor range from $\sim 6-10\%$ in the three galaxies.

To further extend the analysis, we need to find the mass of molecular gas in 
the clouds whose surfaces produce the PDR emission. NGC 1569 is the only 
galaxy of the three for which we have a measured value of the CO luminosity,
$L_{CO}=35\,L_{\sun}$, from Young \et\ (1995). The standard conversion 
factor, $L_{CO}/M_{H_2}\sim 8\times 10^{-6}L_{\sun}/M_{\sun}$ (e.g. Young 
\et\ 1986) has been shown by 
Taylor \et\ (1999) to be too large by a factor of $\sim 6$ in this galaxy,
reflecting a metallicity about 1/6 solar. By applying this correction to the 
standard conversion factor, we find a molecular gas mass M(H$_2$) in NGC 1569 of 
$1.9\times 10^7\,M_{\sun}$, about 1.5 times the atomic gas 
mass. Values of M(H$_2$) are given in Table \ref{tabmodel}.

A comparison of the surface area of the clouds (the [\CII] measurement)
with the volume of the clouds (the CO measurement) allows us to characterize 
the molecular clouds in NGC 1569,
and the derived parameters are given in Table \ref{tabcloud}. 
If we assume that the [\CII] and CO regions
have equal density (that is, $n$(H$_2$)/$n_a$$\sim$1), 
then we find the emission is fit by an ensemble 
of 4 giant molecular clouds with typical radius $\sim 25$ pc and cloud mass 
$\sim 8\times 10^6 M_{\sun}$. If, instead, we assume that the CO region has 
a density 3 times that of the atomic gas ($n$(H$_2$)/$n_a$$\sim$3),
the number of clouds increases to 
8, the mass per cloud decreases by a factor 2, and the cloud radius falls to 
18 pc. These results for number of clouds and cloud
size are similar to those of Taylor \et\ (1999), who 
found from CO observations that the GMCs in NGC 1569 could be resolved into 
$\sim 4-5$ clouds with radii $\sim 20$ pc, though we note that they 
report cloud masses which are significantly lower than ours. 
Our clouds are $\sim$20 times more massive than a typical Milky
Way GMC (Scoville \& Sanders 1987).

The PDR surface layer on our clouds would extend $\sim$6 pc in from 
the cloud surface, a distance larger than would be found if the clouds had 
solar metallicities. Given the derived density of the PDR gas, $n=3.5\times 
10^3\,\rm cm^{-3}$, these GMCs 
have a gas column density from surface to center of $\sim 3\times 10^{23}\,
\rm cm^{-2}$, corresponding to a visual extinction of 
A$_V$$\sim 22$ assuming that 
dust is depleted by the same factor as the carbon. 
As a consistency check, we
compare these results with those of Kaufman \et\ (1999) who find that clouds
with $A_V\sim 10$, $n=10^3\,\rm cm^{-3}$, 
and metallicities of a few tenths 
solar have L(CO)/M(H$_2$) conversion factors of 
$\sim 10^{-6}L_{\odot}/M_{\odot}$, very close to the corrected 
value of Taylor \et\ (1999). 
The results for the NGC 1569 clouds give A$_V$ higher by a factor of 
$\sim 3-4$
than the A$_V$ 
predicted by the photoionization-regulated star formation model of 
McKee (1989). This may imply that the clouds in NGC 1569 have not yet reached
equilibrium, which may be consistent with the disruption of the ISM from
the recent starburst and the proximity of the GMCs to the super star clusters
(Taylor \et\ 1999).
(Note that these high extinction values only apply to the GMCs which
are not necessarily visible in the optical).

One thing that is clear from these observations is that the PDR observations
in irregulars
are sensitive, not to the average density of GMCs, but to the conditions in the
denser clumps within the GMCs. In the Milky Way, the average GMC density is 
of order 100 cm$^{-3}$, but observations of FUV illuminated gas give PDR
densities of order $10^3\,\rm cm^{-3}$ and higher. Such is the case, 
for example, in Orion where the average density of the entire molecular cloud 
complex is far lower than that in the FUV illuminated clumps region at the 
edge of the Orion Nebula. 

A similar analysis to the one carried out above may be done for the other two 
galaxies, NGC 1156 and IC 4662. However, in both cases we only have upper 
limits to the molecular mass. As a result, we can only derive ranges for the 
numbers and masses of GMCs in these galaxies with large uncertainties. 
In NCG 1156, we find between 
1 and 6 GMCs with masses ranging from $\sim 7\times 10^7$ M\solar\ 
to $3\times 10^8$ M\solar;
for IC 4662, we find between 1 and 2 GMCs with masses from 
$\sim 3\times 10^6$ M\solar\ to $8\times 10^6$ M\solar. These GMCs are
also large compared to those in the Milky Way (Scoville \& Sanders
1987): 175--750 times more massive in NGC 1156 and 7--20 times in IC 4662.

\subsubsection{Uncertainties in PDR Modelling}

To quantify the uncertainties in deriving $G_0$ and $n$, we have used
several different methods for each galaxy.
In the PDR models, we typically compare one line ratio 
(usually, L$_{[CII]}$/L$_{[OI]}$)  
with the infrared line-to-continuum ratio, 
(L$_{[OI]+[CII]}$/\ltir). If we do this   
using the raw [\OI], [\CII] and FIR values, we get certain values for $G_0$ and
$n$, but we suspect that much of the [\CII] emission comes from 
ionized gas, not PDRs. Thus, we correct [\CII] to get [\CII]$_{PDR}$. 
The correction
is based on the observed [\NII] flux with the assumption that [\NII] only
comes from ionized gas; given an assumed [\CII]-to-[\NII] abundance ratio, we
can calculate the [\CII] from ionized gas and subtract that portion from the 
total [\CII], leaving only the PDR contribution. In all three galaxies, 
we assumed
that the C/N ratio was 4.6.
However, for all three galaxies
we only have upper limits on the [\NII] flux, so we have a range of possible 
[\CII] from PDRs. 
In one variation on the standard method, if
one assumes that there is no [\NII], then all of the [\CII]
comes from PDRs, and the models give significantly different answers for
$G_0$ and $n$. 
A second variation is to solve for $G_0$ and $n$ using other ratios. 
In this case,
we compared $L_{[OI]}$/\lfir\ with $L_{60}/L_{100}$. 
This has the advantage that we no
longer rely on the [\CII] measurements, though we do have to assume that 
the 60 and 100 micron fluxes are produced in PDRs. Again, we get somewhat
different results for $G_0$ and $n$. 

In Figure \ref{figuncertain}, we summarize the results of the 
three methods. 
These three methods together give an 
estimate of the uncertainty in the model values of
$G_0$ and $n$ and hence clouds masses
and sizes.
One can see that the three methods give answers that can be uncertain 
by up to an 
order of magnitude in $G_0$ and $n$. These uncertainties almost surely
are larger than the uncertainties in the measurements. What is heartening is 
that the preferred method 
gives results within
a factor $\sim$3 of the other two methods.

\section{Summary and Comments}

We have discussed mid-infrared imaging and FIR spectroscopy
of a sample of 5 IBm galaxies observed by {\it ISO} as part
of a larger study of galaxies of all Hubble types.
The galaxies include NGC 1156, NGC 1569, NGC 2366, NGC 6822, and IC 4662.
The galaxies NGC 1156, NGC 1569, and IC 4662 are high luminosity, 
high surface brightness,
and high star formation rate systems relative to median properties of
larger samples of irregulars, and so are not typical irregulars.
NGC 2366 is dominated by a supergiant \HII\ complex, and NGC 6822,
although more of a typical irregular,
yielded useful data with
CAM.

The mid-infrared imaging of all 5 galaxies 
is in two bands: one at 6.75 \mum\ that
is dominated by PAHs and one at 15 \mum\ that is dominated by
small dust grains. The spectroscopy of 3 of the galaxies (NGC 1156, NGC 1569,
and IC 4662)
includes [\CII]$\lambda$158 \mum\
and [\OI]$\lambda$63 \mum, important coolants of PDRs,
and [\OIII]$\lambda$88 \mum\ and [\NII]$\lambda$122 \mum, that
come from ionized gas regions. [\OI]$\lambda$145 \mum\ and [\OIII]$\lambda$52
\mum\ were measured in NGC 1569 as well.
We compare the observations of the irregulars with the larger sample
of spiral galaxies observed as part of our {\it ISO} program.

We observe the following:

\begin{itemize}

\item{In the mid-infrared images most of the emission we detect is
associated with the brightest \HII\ regions in the galaxies,
and the integrated \lwcolor\ ratios are comparable to what is
observed in regions of star formation.}

\item{The ratio of PAH-to-small grain emission drops as the 
FIR color temperature becomes hotter in all galaxies, and the
irregulars in our sample lie mostly
at the warmer \irascolor\ and lower \lwcolor\ end of the distribution.}

\item{The ratio of PAH-to-FIR emission drops as the FIR color
temperature becomes hotter while the small grain emission-to-FIR
ratio remains more nearly constant in all galaxy types.}

\item{The irregular galaxies in our sample 
have low PAH emission at 6.75 \mum\ and
low small grain emission at 15 \mum\ relative to \ha\ emission,
compared to spirals. The \lwtwoha\ ratio drops with increasing
FIR color temperature, while the \lwthreeha\ 
ratio is constant among our irregulars.}

\item{The PAH and small grain emissions relative to \ha\
both increase as \lfir/\lb\ increases for all galaxies,
possibly with a plateau in the value of \flwthree/\fha\
for low values of \lfir/\lb. Our irregular galaxies
lie at the low \lfir/\lb\ end of the distributions.}

\item{The [\CII] emission is higher relative to FIR emission
in these irregulars than in spirals of the same FIR color temperature.
The [\CII] line represents 0.3\% to 1\% of the
FIR emission of the irregulars.}

\item{The [\CII] emission in our irregulars is high
relative to the PAH 6.75 \mum\ emission compared to spirals, 
and \fcii/\flwtwo\ is constant
among the irregulars. 
On the other hand, for small grains, the \fcii/\flwthree\
ratio drops as the FIR color temperature increases for all
galaxies, but the \fcii/\flwthree\ ratio is high
in the irregulars compared to spirals with the same FIR color temperature.}

\item{For our irregulars the ratio of [\OIII] to [\CII] is very high
compared to spirals. In addition, 
the high [\CII] to 6.75 \mum\ and FIR emissions
suggest that [\OIII] is high relative to the infrared continuum
and other lines as well.}

\item{For all galaxies the \fcii/\fha\ ratio increases as
\lfir/\lb\ increases, and our irregulars are at the low
\fcii/\fha, low \lfir/\lb\ end of the distribution.}

\item{In the irregulars in our sample
the [\OIII] emission is high with respect to
[\CII].}

\item{
The L$_{[CII]}$/L$_{CO}$ ratio is very high among our sample of
irregulars.}

\item{From PDR models we derive physical conditions in the PDRs
of NGC 1156, NGC 1569, and IC 4662:
Radiation fields relative to the solar neighborhood $G_0$ 
are $10^{3.0}$--$10^{4.4}$, 
gas densities $n$ are $10^{3.1}$--$10^{4.4}$ cm$^{-3}$,
and pressures $P$ are $10^{5.6}$--$10^{7.3}$ cm$^{-3}$ K.
In NGC 1569, where L$_{CO}$ has been measured,
we deduce the presence of 4 GMCs with masses of that are about
20 times higher than that of a typical GMC in the Milky Way, and 
the PDR is 6 pc thick. 
}

\end{itemize}

The upper limits on the [\NII]$\lambda$122 \mum\ emission in 
our irregulars indicate that a large fraction
of the [\CII]
emission detected in these irregulars comes from PDR gas and the
component from \HII\ regions is small although these conclusions do depend on
physical conditions being similar to spirals. 
Less than 39\% of the [\CII]
originates from \HII\ regions.
In spite of the high [\OIII] and \ha\ emission relative to
[\CII],
the contribution
of ionized gas regions to [\CII] emission is similar to
or less than the fractions observed for spirals. 

The average stellar effective temperatures in NGC 1569 and IC 4662
are high compared to measurements in other galaxies.
In NGC 1569 this is consistent with the presence of two young super
star clusters and luminous \HII\ regions, and suggests an end
to the recent starburst of $\leq$6 Myrs ago.
This result is also consistent with the measurement
of exceptionally high \foiii/\fcii\ ratios in our irregulars. The high
excitation stars ionize most of the \HII\ regions to higher ionization
states, reducing the amount of [\CII] in
\HII\ regions. 
Our irregulars also have higher \lcii/\lfir\ ratios
for their FIR color temperatures relative to spirals. This can
be understood if the spirals have a larger contribution 
to heating of the dust from cooler stars, presumbably from the
general interstellar radiation field, 
than do the irregulars.

The increase in \lwtwoha\ and \lwthreeha\ with increasing \lfir/\lb\
tells us that the PAH and small grain emissions are related to the total
dust emission.
However, the decrease in PAH emission in the irregulars in our sample
relative to small grain,
FIR, and \ha\ emissions for increasing FIR color temperature is interpreted
as a decrease in PAH emission resulting from an increase in the 
radiation field due to star formation, 
either through destruction of
the PAH itself or of the 6.75 \mum\ carrier on the PAH.

That the \lwthreeha\ ratio is constant among the irregulars that
we observed means that the more hot stars that are formed,
the more small grains are heated. 
The 15 \mum\ emission may come primarily from small grains in \HII\ regions.
We interpret the linear dependence of the
15 \mum\ flux with H$\alpha$ to mean that the 15 \mum\ emission
is being generated by the transient heating of small dust grains by
single photon events, possibly
Ly$\alpha$ photons trapped in \HII\ regions. 
The low \lwthreeha\ ratio, as well as the high
\fcii/\flwthree\ ratio, in our irregulars compared to spirals
may be due to the 
lower dust content overall resulting in dust grains being, 
on average, further away
from the heating source. By contrast, it has been suggested that
the 15 \mum\ emission in spirals may be warm dust grains radiating
on the Wien side of the blackbody spectrum rather than transient single
photon events. This interpretation is consistent with the
increase in \lwthreeha\ ratio with hotter FIR color temperature
seen in spirals, but not irregulars.

The high [\CII] emission relative to FIR, PAH emission at 6.75 \mum,
and small grain emission at 15 \mum\ in our irregulars is harder to interpret.
The small scatter in the \fcii/\flwtwo\ ratio among spirals
has been interpreted as meaning that the PAHs that produce
the 6.75 \mum\ emission and the PAHs that heat the PDR and
hence produce the [\CII] emission are the same entities. But, then the much
higher \fcii/\flwtwo\ ratio in our 
irregulars compared to spirals would require that
the PAHs in irregulars produce several times more heat than the PAHs in spirals.
Alternatively, the small scatter in \fcii/\flwtwo\ among spirals could be
explained if the carrier of the 6.75 \mum\ feature tracks
the heating but contributes only a part of the heating, which is due mostly
to small grains or other PAHs. In this scenario these irregulars 
are understood if they have a higher proportion of the PAHs that heat 
the PDRs to the PAHs that produce the 6.75 \mum\ feature.
A different incident spectral energy distribution could also play a role.

Our data give some clues to the geometry of the clouds in the irregulars
in our sample.
The high [\OIII]/[\CII] ratio requires
a small solid angle
of optically thick ($A_V>1$) PDRs outside the \HII\ regions.
That is, if the \HII\ region is surrounded by an optically thick
PDR, the copious production of [\CII] there would lower the
\foiii/\fcii\ ratio. So, the optically thick component of the PDR
must present a smaller covering factor in our irregulars compared to
spirals.
This picture is also consistent with the fact that PAH 6.75 \mum\ 
and [\CII] emissions, both of which come primarily from PDR gas,
are lower compared to \ha.
On the other hand, the L$_{[CII]}$/L$_{CO}$ ratio is very high among our sample
of irregulars, higher than values measured
in spirals or \HII\ regions. 
A picture in which the clouds
in our irregulars have 
very thick [\CII]
shells around tiny CO cores compared to clouds in spirals
is consistent with physical parameters we derive for NGC 1569's PDRs.

From models we deduce the physical conditions in the PDRs.
The FUV stellar radiation field $G_0$ is like that in typical spirals
in NGC 1156 but much higher in the starburst galaxy NGC 1569 and in IC 4662.
That NGC 1569 stands out in $G_0$ makes sense since it is a starburst
system.
We emphasize, however, that the conditions measured by the data presented
here apply to very local conditions in star-forming clouds, particularly
the denser clumps within GMCs. That is why
NGC 1569, in fact, does not appear far more extreme than it does.

Finally, we estimated the number of molecular clouds,
their masses, and their sizes within the LWS beam 
in NGC 1156, NGC 1569, and IC 4662. 
We deduce the presence of a few clouds in each galaxy with
masses much larger than typical Milky Way GMCs. 
These extraordinarily large clouds may also be necessary to form 
super star clusters, as NGC 1569 has recently done.

The irregular galaxies in our sample have star formation rates
normalized to their size that are comparable to those in spirals.
In fact, the star formation rates span the entire range observed
in our spiral sample,
including the high end of the range. However, the infrared
properties of the irregulars are dominated by star-formation, perhaps
more than is the case in spirals. 
This implies that the ISM beyond the \HII\ regions and associated PDRs
in irregulars
is too faint to contribute measureably to the {\it ISO} observations.
We know that the ISM in irregulars can be quite clumpy and that
ionizing photons can travel large distances.
But, the low contribution of the ISM beyond \HII\ regions to the infrared
may be a consequence of a lower stellar radiation field outside
star-forming regions and reflect the general low surface brightness
disk of irregulars. In galaxies with low dust column densities,
without concentrations of hot stars to 
heat the dust and gas, the ISM becomes very hard to detect
in PAH, small grain, and PDR emission features.

What does this imply about more normal irregulars? 
Most of the irregular galaxies in our sample are not representative of
typical irregular galaxies. Several in our sample are very high
surface brightness and one is a starburst. Most irregulars
are lower in surface brightness and star formation activity.
Unfortunately, we were not able to observe these galaxies with {\it ISO}.
However, it is likely that more typical irregulars, if they could
be observed, would also turn out to be dominated by star-forming
regions since these would be the primary contributors to the FIR
and mid-infrared emission. The irregulars can be seen as 
containing glowing islands of star-forming
regions in a sea of otherwise relatively dim gas and stars.

\acknowledgments

We are grateful to P.\ Massey and C.\ Claver
for the use of their optical images. 
This work was supported in part by {\it ISO} data analysis funding from the 
US National Aeronautics and Space Administration through the Jet Propulsion
Laboratory (JPL) of the California Institute of Technology and in part by
support to D.A.H. from the Lowell Research
Fund and grant AST-9802193 from the National Science
Foundation.
The \ha\ imaging would not have been possible without filters purchased through
funds provided by a Small Research Grant from the American Astronomical Society,
National Science Foundation grant AST-9022046,
and grant 960355 from JPL.
D.A.H. wishes to thank Lowell Observatory for the observing time on the
Perkins 1.8 m telescope that produced the northern \ha\ images
and optical long-slit spectroscopy.

\begin{table}
\dummytable\label{tabgal}
\end{table}

\begin{table}
\dummytable\label{tabisoobs}
\end{table}

\begin{table}
\dummytable\label{tabcam}
\end{table}

\begin{table}
\dummytable\label{tablws}
\end{table}

\begin{table}
\dummytable\label{tabiras}
\end{table}

\begin{table}
\dummytable\label{tabmodel}
\end{table}
 
\begin{table}
\dummytable\label{tabcloud}
\end{table}
 
\clearpage

\clearpage

\figcaption{\protect\ha\ images of the galaxies in our
sample shown with V-band contours superposed. 
The stellar continuum has been subtracted from the \protect\ha\ images.
The stars have been removed from the V-band images as well for
NGC 1156 and NGC 2366, and the V-band images of NGC 2366 and NGC 6822
were smoothed before contouring.
The V-band images of NGC 1156, NGC 1569, and NGC 2366 are courtesy of
P.\ Massey, and the V-band image of NGC 6822 is courtesy of C.\ Claver.
The image of IC 4662 shows a region, labeled IC 4662-A, 
that is well separated from the
main body of the galaxy (890 pc at IC 4662). 
It has the H$\alpha$ luminosity of a large
OB association. Three tiny \protect\HII\ regions between IC 4662
and IC 4662-A have \protect\ha\ luminosities similar to that of
the Orion nebula.
One arcminute corresponds to 2.3 kpc at NGC 1156, 0.7 kpc at NGC 1569,
1.0 kpc at NGC 2366, 0.14 kpc at NGC 6822, and 0.6 kpc at IC 4662.
\label{figvha}}

\figcaption{Spectra of FIR emission lines observed with
{\it ISO}. The y-axes are in units of 
10$^{-14}$ W cm$^{-2}$ $\mu$m$^{-1}$.
The spectra of IC 4662 include two observations of each line.
The two observations differ markedly in the continuum levels
due to calibration uncertainties in the dark current subtraction.
\label{figlws}}

\figcaption{Mid-infrared CAM images of the irregular galaxies.
The LW2 filter was centered at 6.75 \protect\mum.
Objects labeled ``star'' are
foreground stars, and the object in NGC 2366 labeled ``Gal''
is a background galaxy (Drissen et al.\ 2000).
Contours are of \protect\ha\ emission
where the \protect\ha\ images have been transformed to the 
geometry of the CAM images and pixel scale.
Named \protect\HII\ regions are labeled in NGC 2366 and
NGC 6822 (Hodge 1977, Killen \& Dufour 1982).
In the image of NGC 1569, the four plus signs mark the positions of
the 5 giant molecular clouds mapped by Taylor et al.\ (1999),
where two of the clouds are given as one position,
and the two five-pointed stars mark the positions of the super 
star clusters.
One arcminute corresponds to 2.3 kpc at NGC 1156, 0.7 kpc at NGC 1569,
1.0 kpc at NGC 2366, 0.14 kpc at NGC 6822, and 0.6 kpc at IC 4662.
\protect\ha\ contours, in units of $10^{34}$ ergs s$^{-1}$ arcsec$^{-2}$,
are 7.8 and 46.9 for NGC 1156; 1.1, 5.4, and 42.8 for NGC 1569;
0.2, 1.4, and 6.6 for NGC 2366; 3.1$\times10^3$ for NGC 6822;
and 0.2, 0.4, 0.6, 2.4, and 11.4 for IC 4662.
\label{figcam}}

\figcaption{Mid-infrared ratio \protect\lwcolor\ plotted
against the FIR ratio \protect\irascolor\ and
the area-normalized star formation rate (SFR). The dashed line
marks a \protect\lwcolor\ of 1. The area used to normalize the
star formation rate is $\pi$R$_{25}^2$.
\label{figlwcolor}}

\figcaption{Ratio of mid-infrared flux to \protect\ha\
flux for LW2 (PAH-dominated) and LW3 (small dust grain dominated). 
This is plotted against the 
FIR ratio \protect\irascolor\ and the
ratio of the integrated FIR flux to the B-band flux.
The \protect\lha\ have been corrected for reddening.
\label{figlwhalpha}}

\figcaption{Ratio of mid-infrared flux to FIR 
flux for LW2 (PAH-dominated) and LW3 (small dust grain dominated). 
This is plotted against the 
FIR color ratio \protect\irascolor.
\label{figlwfir}} 

\figcaption{Ratio of [\protect\CII]$\lambda$158 \protect\mum\
emission to mid-infrared LW2 and LW3 emission.
These are plotted against the FIR ratio
\protect\irascolor\ and the ratio of the integrated FIR
to the B-band flux of the galaxy.
The f$_{6.75}$ and f$_{15}$ are in units of ergs cm$^{-2}$ s$^{-1}$
Hz$^{-1}$ and the units of [\protect\CII] are ergs cm$^{-2}$ s$^{-1}$,
so the ratios have units of Hz$^{-1}$.
\label{figlwcii}}

\figcaption{Ratio of [\protect\OIII]$\lambda$88 $\mu$m to [\protect\CII] 
emission plotted against
\protect\irascolor.
\label{figoiiicii}}

\figcaption{Ratio of \protect\fciifir\ plotted against 
the \protect\irascolor\ and
star formation rate per unit area.
Data for the LMC comes from Israel et al.\
(1996), Kennicutt \& Hodge (1986), RC3, and Rice et al.\ (1988).
\label{figciifir}}

\figcaption{Ratio of [\protect\OI]$\lambda$145 \protect\mum\ to [\protect\OI]$\lambda$63
\protect\mum\ versus L$_{FIR}$/L$_B$ for spirals in the distant galaxy
sample (dots) and for the irregular NGC 1569 (star).
\label{figoioi}}

\figcaption{Ratio of [\protect\CII]$\lambda$158 \protect\mum\ 
to \protect\lha\
plotted against the L$_{FIR}$/L$_B$ ratio.
\protect\lha\ has been corrected for reddening.
\label{figciiha}}

\figcaption{Ratio of [\protect\OI]$\lambda$63 \protect\mum\ to [\protect\CII]$\lambda$158
\protect\mum\ plotted against the FIR color.
In the top panel [\protect\CII] is the total measured. In the bottom panel
an estimate of the contribution of \protect\HII\ regions to [\protect\CII] has been subtracted
to leave only emission from PDRs. For the three irregulars NGC 1156, NGC
1569, and IC 4662 the corrections assume the full value of
upper limits.
For IC 10 and the distant
galaxy sample the corrections
were determined from [\protect\NII]$\lambda$122 \protect\mum\ as described
by Kaufman et al.\ (1999).
\label{figoicii}}

\figcaption{
Derived values of $G_0$ and $n$ for the three irregular galaxies NGC 1156, 
NGC 1569 and IC 4662. Squares show the values of $G_0$ and $n$ determined
using uncorrected values of the [\protect\CII] intensity in 
the comparison of L$_{[OI]}$/L$_{[CII]}$
with L$_{[OI]+[CII]}$/\protect\ltir; 
triangles show the values determined when 
the value of [\protect\CII] corrected for the \protect\HII\ region
contribution is used in the same comparison; 
and pentagons show the values
determined when L$_{[OI]}$/\protect\ltir\ is compared with the IRAS 
\protect\irascolor\ ratio. The
third method is independent of the [\protect\CII] correction (see text).
The results from these three methods give the reader a feel for the
uncertainties in the modelling of PDR properties.
\label{figuncertain}}

\end{document}


\begin{deluxetable}{lrrr}
\tablenum{4}
\tablecaption{Line Fluxes and Ratios\tablenotemark{a}. \label{tablws}}
\tablewidth{0pt}
\tablecolumns{4}
\tablehead{
\colhead{Quantity} & \colhead{NGC 1156} & \colhead{NGC 1569}
& \colhead{IC 4662} 
} 
\startdata
f$_{[OIII]\lambda52}$\tablenotemark{b} ~\dotfill       & \nodata   & 1.853     & \nodata   \nl
                             & \nodata   & 0.104     & \nodata   \nl
f$_{[NIII]\lambda57}$\tablenotemark{b} ~\dotfill       & \nodata   & $\leq$0.206 & \nodata \nl
f$_{[OI]\lambda63}$\tablenotemark{b} ~\dotfill         & 0.082     & 0.659     & 0.145     \nl
                             & 0.006     & 0.051     & 0.011     \nl
f$_{[OIII]\lambda88}$\tablenotemark{b} ~\dotfill       & \nodata   & 2.768     & 0.458     \nl
                             & \nodata   & 0.091     & 0.017     \nl
f$_{[NII]\lambda122}$\tablenotemark{b} ~\dotfill       & $<$0.0084 & $<$0.0319 & $<$0.0107 \nl
f$_{[OI]\lambda145}$\tablenotemark{b} ~\dotfill        & \nodata   & 0.0134    & \nodata   \nl
                             & \nodata   & 0.0025    & \nodata   \nl
f$_{[CII]\lambda158}$\tablenotemark{b} ~\dotfill       & 0.156     & 0.599     & 0.125     \nl
                             & 0.010     & 0.020     & 0.006     \nl
\protect\lcii/\protect\lfir\tablenotemark{c} ~\dotfill & 0.0051    & 0.0025    & 0.0027    \nl
                                             & 0.0012    & 0.0004    & 0.0001    \nl
\protect\lcii/\protect\lha\tablenotemark{d} ~\dotfill  & 0.18      & 0.07      & 0.06      \nl
                                             & 0.02      & 0.00      & 0.00      \nl
\protect\lha(LWS beam)/\protect\lha(total)  ~\dotfill  & 0.70      & 0.76      & 0.79      \nl
\protect\lcii/\protect\lco\tablenotemark{e}  ~\dotfill & $>$10000  & 37000     & $>$34000  \nl
CO beam area/LWS beam area   ~\dotfill                 & 0.64      & 0.36      & 0.33      \nl
log \protect\fcii/f$_{6.75}$\tablenotemark{f} ~\dotfill & 12.25     & 12.27     & 12.40     \nl
                                             &  0.09     &  0.08     &  0.09     \nl
log \protect\fcii/f$_{15}$\tablenotemark{f} ~\dotfill  & 12.04     & 11.62     & 11.88     \nl
                                             &  0.09     &  0.09     &  0.09     \nl
f$_{[OIII]\lambda52}$/f$_{[OIII]\lambda88}$ ~\dotfill & \nodata   & 0.67      & \nodata   \nl
                                             & \nodata   & 0.04      & \nodata   \nl
$n_e$ (cm$^{-3}$)     ~\dotfill                        & \nodata   & $<$300        & \nodata   \nl
f$_{[OIII]\lambda88}$/f$_{[CII]}$  ~\dotfill           & \nodata   & 4.61      & 3.66      \nl
                                             & \nodata   & 0.18      & 0.23      \nl
f$_{[OI]\lambda63}$/f$_{[OI]\lambda145}$  ~\dotfill    & \nodata   & 49.2      & \nodata   \nl
                                             & \nodata   &  9.9      & \nodata   \nl
L$_{[OI]\lambda63}$/\protect\lcii   ~\dotfill          & 0.39      & 1.10      & 1.16      \nl
                                             & 0.22      & 0.08      & 0.07      \nl
L$_{[CII],PDR}$/\protect\lcii   ~\dotfill              & $\geq$0.76 & $\geq$0.76 & $\geq$0.61 \nl
L$_{[OI]\lambda63}$/L$_{[CII],PDR}$\tablenotemark{g}  ~\dotfill 
                                             & 0.53--0.69  & 1.10--1.44 & 1.16--1.89 \nl
\tablebreak
L$_{H\alpha}$/L$_{[OIII]\lambda88}$\tablenotemark{h} ~\dotfill & \nodata & 2.48 & 2.07 \nl
                                                     & \nodata & 0.46 & 0.40 \nl
\enddata
\tablenotetext{a}{Second, unlabelled line contains the uncertainties in the values
in the first line.}
\tablenotetext{b}{Units of 10$^{-18}$ W cm$^{-2}$.}
\tablenotetext{c}{The far-infrared luminosity is the global 
galactic value, not just that contained within the LWS aperture.}
\tablenotetext{d}{The H$\alpha$ luminosity is the global 
galactic value.}
\tablenotetext{e}{The CO fluxes are not corrected for the 
differing beam size compared to the LWS aperture. The ratio of the beam sizes
is given in the next line.
CO upper limits are taken from Hunter \& Sage (1993) and Heydari-Malayeri
\et\ (1990), assuming a line width of 20 \protect\kms\ and a conversion
factor of 34 to go from K \protect\kms\ to Jy \protect\kms.
The CO value for NGC 1569 is from Young \et\ (1995) without their
deconvolution with an exponential disk model.
Conversion to L$_{CO}$ uses the formula of Kenney \& Young (1989), 
that uses a Galactic-derived conversion factor.}
\tablenotetext{f}{f$_{6.75}$ and f$_{15}$ are in units of 
ergs cm$^{-2}$ s$^{-1}$ Hz$^{-1}$ and the units of [CII] are 
ergs cm$^{-2}$ s$^{-1}$, so the ratio has units of Hz$^{-1}$.}
\tablenotetext{g}{The lower end of the range corresponds to no correction for
the contribution of \protect\HII\ regions to the [CII] emission;
the upper end corresponds to the total amount of the upper limit
of the correction given above.}
\tablenotetext{h}{L$_{H\alpha}$ is the H$\alpha$ luminosity in the LWS
beam.}
\end{deluxetable}